\pdfoutput=1
\documentclass[10pt]{iopart}

\usepackage[pdftex]{graphicx}
\usepackage[labelformat=empty,position=top]{subcaption}
\usepackage[export]{adjustbox}
\usepackage{floatrow}
\usepackage{color}
\usepackage{mathrsfs}
\usepackage[english]{babel}
\usepackage{epsfig}
\usepackage{epstopdf}
\usepackage{babelbib}
\usepackage{bm}
\usepackage{tcolorbox}
\usepackage{tabularx}
\usepackage{array}
\usepackage{colortbl}
\tcbuselibrary{skins}
\usepackage{comment}
\usepackage{longtable}
\usepackage{iopams}
\usepackage{xr-hyper}
\usepackage{hyperref}
\usepackage[final]{pdfpages}

\begin{document}

\title[Varieties of charge distributions in viral coat proteins]{Varieties of charge distributions in coat proteins of ssRNA+ viruses}

\author{An\v{z}e {Lo\v{s}dorfer Bo\v{z}i\v{c}}$^1$ and Rudolf Podgornik$^{1,2}$}
\address{$^1$ Department of Theoretical Physics, Jo\v zef Stefan Institute, SI-1000 Ljubljana, Slovenia}
\address{$^2$ Department of Physics, Faculty of Mathematics and Physics, University of Ljubljana, SI-1000 Ljubljana, Slovenia}
\ead{anze.bozic@ijs.si}

\begin{abstract}
A major part of the interactions involved in the assembly and stability of icosahedral, positive-sense single-stranded RNA (ssRNA+) viruses is {\em electrostatic in nature}, as can be inferred from the strong $pH$- and salt-dependence of their assembly phase diagrams. Electrostatic interactions do not act only between the capsid coat proteins (CPs), but just as often provide a significant contribution to the interactions of the CPs with the genomic RNA, mediated to a large extent by positively charged, flexible N-terminal tails of the CPs. In this work, we provide two clear and complementary definitions of an N-terminal tail of a protein, and use them to extract the tail sequences of a large number of CPs of ssRNA+ viruses. We examine the $pH$-dependent interplay of charge on both tails and CPs alike, and show that -- in contrast to the charge on the CPs -- the net positive charge on the N-tails persists even to very basic $pH$ values. In addition, we note a limit to the length of the wild-type genomes of those viruses which utilize positively charged tails, when compared to viruses without charged tails and similar capsid size. At the same time, we observe no clear connection between the charge on the N-tails and the genome lengths of the viruses included in our study.
\end{abstract}

%
\vspace{2pc}
%
\submitto{\JPCM}
%
%
\ioptwocol

\section{Introduction}

The nature of viral genomes plays a significant role in their encapsidation into fully-formed virions. On the one hand, the double-stranded DNA (dsDNA) or RNA (dsRNA) genomes of numerous viruses need to be packaged into pre-formed capsids by virtue of a strong molecular motor, due to the high charge density and the rigid molecular conformation of the double-stranded nucleic acids on the nanoscale~\cite{Perlmutter2015,Siber2012,Sun2010}. On the other hand, viruses with more flexible single-stranded RNA (ssRNA) genomes tend to spontaneously self-assemble around the RNA filament, where the binding of the capsid coat proteins (CPs) can be guided both by structure- or sequence-specific as well as non-specific interactions~\cite{Perlmutter2015,Siber2012,Ni2013,Rao2006,Siber2008,Schneemann2006}.

In general, the assembly of CPs around an RNA molecule can be a highly-specific process, guided by, e.g., RNA packaging signals (PSs)~\cite{Perlmutter2015,Rao2006}. The presence of PSs alone does not, however, guarantee the packaging of RNA into virions. In addition to specific, localized structural features, RNA secondary and tertiary structure can be of importance for the interaction of the RNA with CPs prior to packaging~\cite{Beren2017,Erdemci2016,Erdemci2014,Yoffe2008,Tubiana2015}. What is more, the length of the RNA molecule itself is an important factor in the assembly, as it naturally carries a significant negative charge. The total charge of RNAs packaged into capsids of different ssRNA viruses is consistently greater than the positive charge of the basic amino acid (AA) residues lining the interiors of the capsids, making these viruses {\em negatively overcharged}~\cite{Perlmutter2015,Siber2012,Garmann2014}. This furthermore implies a direct relationship between the genome length and capsid charge, and in support of the importance of non-specific electrostatics driving RNA encapsidaton, the total positive charge on the capsid inner surface was observed to correlate with the length of the genomic RNA for a diverse group of ssRNA viruses~\cite{Perlmutter2015,Belyi2006,Hu2008,Ting2011}.

The largest contribution to the non-specific electrostatic interactions between CPs and RNA is due to positively charged CP tail groups, whose affinity for RNA varies inversely with the ionic strength of the solution~\cite{Perlmutter2015,Schneemann2006,Bruinsma2016}. These tail groups are extended, highly flexible N-terminal arms and unstructured regions of varying lengths, and are present in the majority of non-enveloped spherical ssRNA+ viruses, while the CPs of enveloped viruses possess only the unstructured regions, with no extended N-terminal arms~\cite{Ni2013,Ni2012}. CP tail groups do not simply play crucial structural roles, but are highly involved in a wide range of biological functions: their disordered nature is essential in promoting correct particle assembly and RNA encapsidation~\cite{Xue2014}, they help in the switching of the CP conformation during assembly~\cite{Ni2013}, and when these tail groups are rich in positively charged residues, they can also control the size of the assembled particles, while their removal can prevent native capsid assembly~\cite{Rao2006}. Interestingly, not all charged tail groups necessarily serve the same function -- experimental evidence shows, for instance, that the tails of Brome Mosaic Virus (BMV) and Cowpea Chlorotic Mottle Virus (CCMV) are not functionally analogous with regard to RNA packaging and seem to employ two distinct packaging mechanisms~\cite{Rao2006}.

Thus, the CP structure of ssRNA viruses can be in general characterized by the presence of two structurally distinct regions: a globular and ordered C-terminal domain involved in the formation of two anti-parallel, four-stranded $\beta$-sheets with a jellyroll topology, and an extended, flexible N-terminal domain that is only partially ordered and thus not observable in the electron density~\cite{Xue2014}. These two structural regions lead to a variety of different interactions during virion assembly, including repulsive CP-CP electrostatic interactions (which inhibit capsid assembly) competing with highly directional, specific CP-CP pairing interactions, and both sequence-specific as well as non-specific electrostatic RNA-CP interactions, which can additionally help to overcome assembly barriers~\cite{Bruinsma2016,Ford2013}. Both experimental and computational observations show that weak interactions are in general required for productive capsid assembly~\cite{Perlmutter2015}, and that conformational flexibility together with the presence of disordered regions -- unusually common in RNA viral proteins -- can be related to the ability to interact with multiple and varying partners~\cite{Tokuriki2009}. A computational study by Perlmutter and Hagan~\cite{Perlmutter2015b} has found that while PSs can confer arbitrarily high specificity of assembly over RNAs with uniform non-specific interactions, the degree of this specificity is overall insensitive to the underlying assembly driving forces, which can be, however, straightforwardly tuned by solution conditions (ionic strength, $pH$) and charge on the CP tail groups. In addition, the specificity conferred by the PSs can lead to kinetic traps in some regions of parameter space, while in others they indeed oversee a highly specific assembly. Their study found that specificity is maximal under conditions where non-specific interactions alone are slightly too weak to promote effective assembly.

\begin{figure*}[!htb]
\begin{center}
\includegraphics[width=0.75\textwidth]{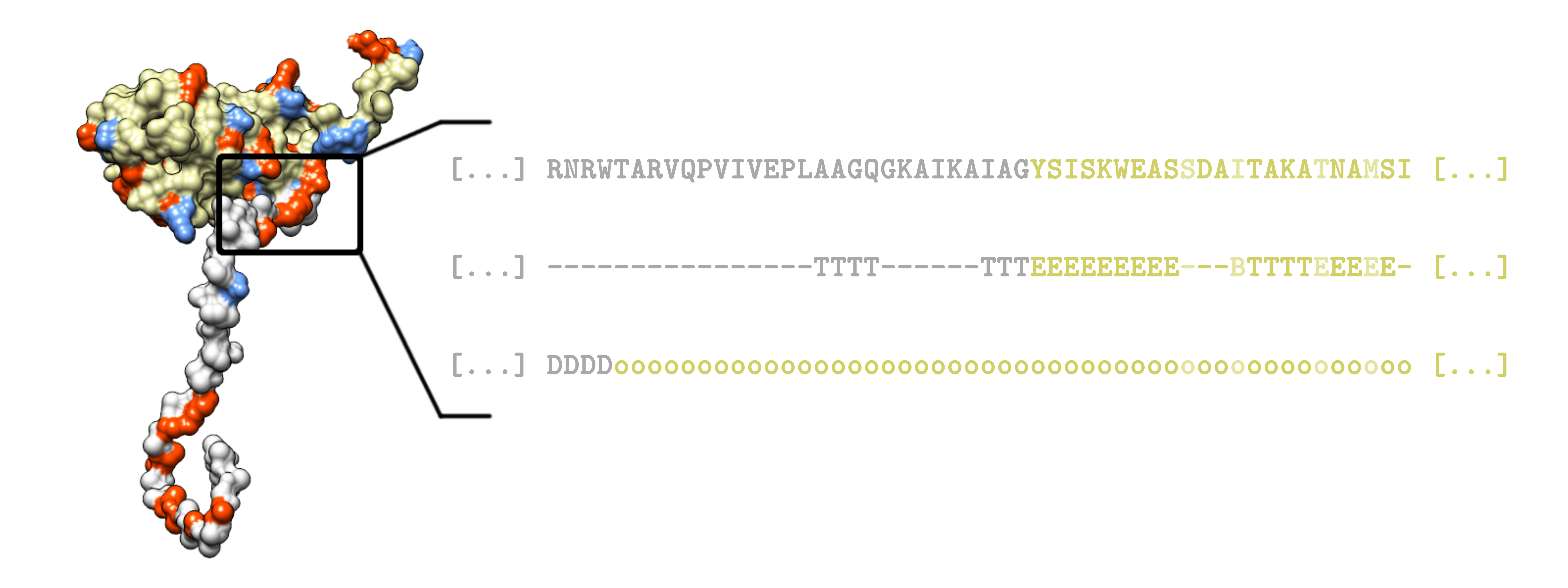}
\end{center}
\caption{Sketch of the capsid protein model. Experimentally determined structure of a viral CP is obtained from PDB/VIPERdb -- in this case, chain C of the CP of BMV (PDB: 1JS9). Ionizable AA residues are shown in red and blue (positively and negatively charged, respectively), while the predicted split of the protein into the N-tail and the structured part of the CP, based on the assigned secondary structure, is shown in gray and beige, respectively. Inset shows the primary AA sequence, assigned secondary structure, and the prediction of intrinsic disorder in a region of the protein. Secondary structure is assigned to the AA residues in the CP using STRIDE, and we define the end of the N-tail as the first occurrence of a given structural element -- in this case, a $\beta$-sheet (E). Alternatively, we split the protein into the N-tail and CP based on the predicted disordered regions (D) in it, where the tail ends at the first occurrence of an ordered region (o) -- in this case, this results in a shorter predicted tail. Afterwards, we assign the ionizable residues in the protein a fractional charge, allowing us to obtain the charge on the tail and the CP at any $pH$. Residues which are predicted to be buried (not exposed to the solvent) are not considered as ionizable, and are highlighted in a lighter color in the primary sequence. 
\label{fig:1}}
\end{figure*}

Non-specific, sequence-independent interactions between RNA and the charged N-terminal tails of CPs are thus predominantly electrostatic in nature and stem from clusters of positive charge, often in form of pronounced arginine-rich motifs (ARMs)~\cite{Perlmutter2015,Schneemann2006,Ni2012}. The fundamental importance of ARMs for RNA-CP interaction in several plant virus genera is well-established, and the positively charged N-tails can be thus envisioned to stabilize encapsidated RNAs within the virus particle~\cite{Rao2006,Perlmutter2013}. The existing works on N-terminal tails of virus CPs have used varying definitions of the N-tails and have either reported a direct correlation between the ssRNA genome length (and thus its total negative charge) and the net charge on the peptide arms~\cite{Belyi2006,Hu2008}, or have concluded that there is no universal genome-to-tail charge ratio~\cite{Ting2011}. In this work, we provide two explicit and complementary definitions of the N-terminal tails -- based either on the secondary structure of the viral CPs or on the presence of intrinsically disordered regions in them -- and use them to classify the N-tails of a large number of ssRNA+ viruses. We compare how the predicted length and charge of the N-tails vary with other parameters, such as genome length and capsid size. In addition, we determine the solvent-accessible surface of the remaining (structured) parts of CPs and obtain the ionizable amino acid residues on them. With this, we are able to study the full $pH$ dependence of the charge on both the N-tails and the CPs, a dependence which has been observed to have a significant impact on the self-assembly of viruses~\cite{Perlmutter2015b,Lavelle2009,Comas2014,Garmann2015,Wilts2015}.

\section{Methods}

\subsection{Dataset of viral coat proteins}

We perform our study on the CPs of various ssRNA+ viruses whose coordinate files are obtained from PDB~\cite{PDB} and VIPERdb~\cite{VIPERdb} databases. In addition, we use the NCBI Nucleotide database~\cite{NCBI} to extract the (approximate) genome lengths of the viruses in our dataset. In a few cases, we also use the UniProt database~\cite{UNIPROT} to obtain full primary sequences of CPs that have incomplete entries in PDB/VIPERdb. While we are chiefly interested in unique viral entries, we also include a few examples of several deposited entries of the same viral CP, in order to estimate the errors of our predictions. In total, our analysis includes $116$ different PDB/VIPERdb entries corresponding to $80$ distinct viruses, which in turn belong to $12$ different viral families: Bromoviridae, Caliciviridae, Dicistroviridae, Flaviviridae, Hepeviridae, Iflaviridae, Leviviridae, Nodaviridae, Picornaviridae, Secoviridae, Tombusviridae, and Tymoviridae; included are also several entries belonging to the genus Sobemovirus, yet unassigned to a family, and several entries of satellite viruses. A list of all the viruses in our dataset is given in Table~\ref{tab:S1a} in the Supplementary Material.

While CP-CP interactions play a major role in virion assembly, their contribution relative to RNA-CP interactions varies among different viruses~\cite{Rao2006,Ni2012,Ni2013}. Throughout this paper, we will present separately viruses from families which are known to utilize positively charged tails (Bromoviridae, Nodaviridae, Sobemovirus, and Tombusviridae), and viruses from those families which do not utilize them. In addition, we will present separately the satellite viruses as well as the viruses belonging to Leviviridae, as the CPs of the latter have a unique fold among non-enveloped ssRNA viruses, where, in the absence of N-terminal tails, the $\beta$-sheet is responsible for interaction with the viral RNA~\cite{Ni2013}.

Majority of the viruses in our dataset have a triangulation number of $\mathcal{T}=3$, meaning that their capsids are composed of $60\mathcal{T}=180$ copies of the same protein. As a consequence of experimental methods and capsid structure reconstruction, the database entries of these viruses usually include $3$ copies of the same CP (the asymmetric unit of the capsid~\cite{VIPERdb}), whose structure can be resolved to a different extent. Our dataset also includes numerous viruses which have a pseudo-$\mathcal{T}=3$ ($\mathcal{T}=p3$) number, their capsids possessing the same symmetry as $\mathcal{T}=3$ capsids, yet consisting of $3$ different proteins. In addition, the asymmetric units of $\mathcal{T}=p3$ viruses  often consist of $4$ and not only $3$ capsid proteins, such as is the case in viruses belonging to Picornaviridae: The first $3$ CPs form the outer surface of the capsid, while the inner surface consists of CP4 and the N-terminal part of CP1~\cite{VIRALZONE}. Taking this into account, we will always calculate the quantities such as N-tail charge as {\em charge per tail}, averaged over the different chains in a dataset entry. In the case of viruses belonging to Nodaviridae, the CP is cleaved in two during virion maturation, and their dataset entries often possess $3$ copies of each of the resulting $2$ proteins. However, during assembly of the CPs and genome into a provirion, where the N-terminal part of the CP plays an important role~\cite{Ni2013,Marshall2001}, the CP is uncleaved~\cite{Odegard2010}. And since the smaller (C-terminal) part of the cleaved protein, deposited in the database, could significantly skew the estimate of charge on the N-tail of the larger part of the protein, we thus exclude the smaller protein from our dataset. Table~\ref{tab:S2a} in the Supplementary Material lists the number of chains used for each database entry, along with the triangulation number of each virus.

\subsection{Protein secondary structure and N-tails}

One of the main aims of this work is to provide a clear and consistent definition of a protein N-tail and analyze its consequences. The N-tail groups of CPs are structurally flexible regions, likely to be intrinsically disordered; such regions are often characterized by a high content of polar and charged residues and a low content of bulky hydrophobic residues~\cite{Ni2013,Xue2014}. They may be ordered in the capsid via interactions with other viral components, but at the same time it is evident that they are flexible in the isolated protein. Taking this into account, we consider two complementary definitions and define the N-tail either
\begin{itemize}
\item[\em (i)] as the part of the CP extending from its N-terminus to the first occurrence of a given element of secondary structure; or 
\item[\em (ii)] as the first intrinsically disordered, contiguous region of the CP, starting again from its N-terminus. 
\end{itemize}
The first definition is the more common one, taking into account the flexibility of the N-tails and contrasting it with the structured part of the CP. On the other hand, the second definition will help us examine the role of disorder in the N-tail regions of viral CPs.

In the first case, {\em (i)}, we use the atomic coordinates of the viral CPs and assign each AA residue a secondary structure using STRIDE~\cite{STRIDE}. (The results obtained using the assignment given by DSSP~\cite{DSSP} match those obtained using STRIDE, and we thus focus only on the latter.) The single-code classification of protein secondary structure given by DSSP/STRIDE involves seven structural elements, such as $3_{10}$ (G), $\alpha$ (H), and $\pi$ (I) helices, hydrogen bonded turns (T), $\beta$-sheets (E), isolated $\beta$-bridges (B), and bends (S). Not all of these elements, however, necessarily correspond to the structurally-ordered part of a protein and could be present in the disordered region as well. Thus, in our definition of a protein N-tail, we terminate it at the first occurrence of any of the following structures: G, H, I, and E. We allow for the presence of T, B, and S structural elements in the tail region, as these represent very short stretches of bonding patterns and should thus not inhibit the flexibility of the N-tail. This definition turns out to be the most consistent in comparisons between individual chains of the same CP, as well as to provide a good match with the predictions of our second definition of an N-tail.

Nevertheless, the above choice of the structural elements signaling the start of a structurally-ordered part of a CP is clearly not the only possible one. What is more, common secondary structure assignment methods can underpredict certain structural elements, such as $\pi$-helices~\cite{Fodje2002}. For this reason, we use a second, {\em (ii)}, independent definition of N-tails and compare it to the first one. We base this definition on the predicted intrinsic disorder in the viral CPs, so that the flexible N-tails should, in general, correspond to a disordered N-terminal stretch of the CP. To predict the intrinsically disordered regions, we use the Metadisorder server (MD2)~\cite{Kozlowski2012}, one of the best predictors of protein disorder, which combines a number of different disorder predictors into a more accurate meta-prediction method. Using MD2, we thus obtain a prediction for the intrinsically disordered parts in the viral CPs based on their AA sequences. For simplicity, we consider only the first contiguous disordered region of the CP starting at its N-terminus to be the N-tail of the protein, which should be a valid assumption in most of the cases.

Due to the flexible, disordered nature of the N-tails, they often remain unresolved in structural experiments, and are usually incomplete in the data deposited in the PDB/VIPERdb databases. To remedy this, we compare the structurally-resolved part of a viral CP with its full primary sequence in order to obtain any missing residues. Such residues are consequently assigned a lack of secondary structure (C or ``--'' in DSSP notation) as well as full disorder (D) for use in the two different N-tail definitions, respectively.

Both N-tail definitions, {\em (i)} and {\em (ii)}, yield in the end an N-terminal sequence of AA residues belonging to the flexible, disordered region of a viral CP. The remaining AA sequence and its assigned secondary structure we then attribute to the structurally-ordered body of the CP. In the rest of the paper, we will refer to the latter region simply as CP, and we will explicitly specify when we will be referring to the entire coat protein including its N-tail. 

\subsection{pH dependence of charges}

To obtain the charges on the CPs and the N-tails at any given $pH$, we follow the procedure fully elaborated previously in Ref.~\cite{ALB2017}. The AA residues we consider as ionizable are the aspartic acid (ASP), glutamic acid (GLU), tyrosine (TYR), arginine (ARG), lysine (LYS), and histidine (HIS). We include the charge on the N- and C-terminus, but we do not consider the acidity of cysteine, a very weak acid which can form disulfide bonds, the exclusion of which should have no qualitative influence on our study~\cite{ALB2017,Nap2014}. The charge on the ionizable residues at a given $pH$ is given by virtue of the Henderson-Hasselbalch equation, which yields the fractional charge of a residue $k$ given its static dissociation constant $pK_a^{(k)}$:
\begin{equation}
\label{eq:hh}
q^{\pm}_k=\frac{\pm1}{1+e^{\pm\ln10(pH-pK_a^{(k)})}}
\end{equation}
for bases ($q^{+}_k>0$) and acids ($q^{-}_k<0$), respectively. For the $pK_a$ values of the different ionizable residues we use the canonical values for isolated AAs~\cite{Nap2014}. Equation~(\ref{eq:hh}) furthermore assumes the limit of relatively high physiological salt concentration, where the electrostatic potential does not induce a significant local shift in $pK_a$ and can thus be ignored~\cite{ALB2017}. In addition, we treat as ionizable only those AA residues which are solvent-accessible. We use STRIDE to determine the relative solvent accessibility (RSA) of each residue, with the cutoff of $c=0.2$ defining the residue accessibility (i.e., $\mathrm{RSA}\geq 0.2$ defining the solvent-accessible residues). As mentioned before, certain parts of the CPs are structurally unresolved and absent in the data -- these residues belong mostly to the flexible N-terminal parts of the proteins. For this reason, we treat any residues missing in the structural data as being completely accessible, assigning to them an $\mathrm{RSA}=1$.

\section{Results}

\subsection{pH dependence of N-tail and CP charge}

\begin{figure}[!t]
\begin{center}
\includegraphics[width=0.95\columnwidth]{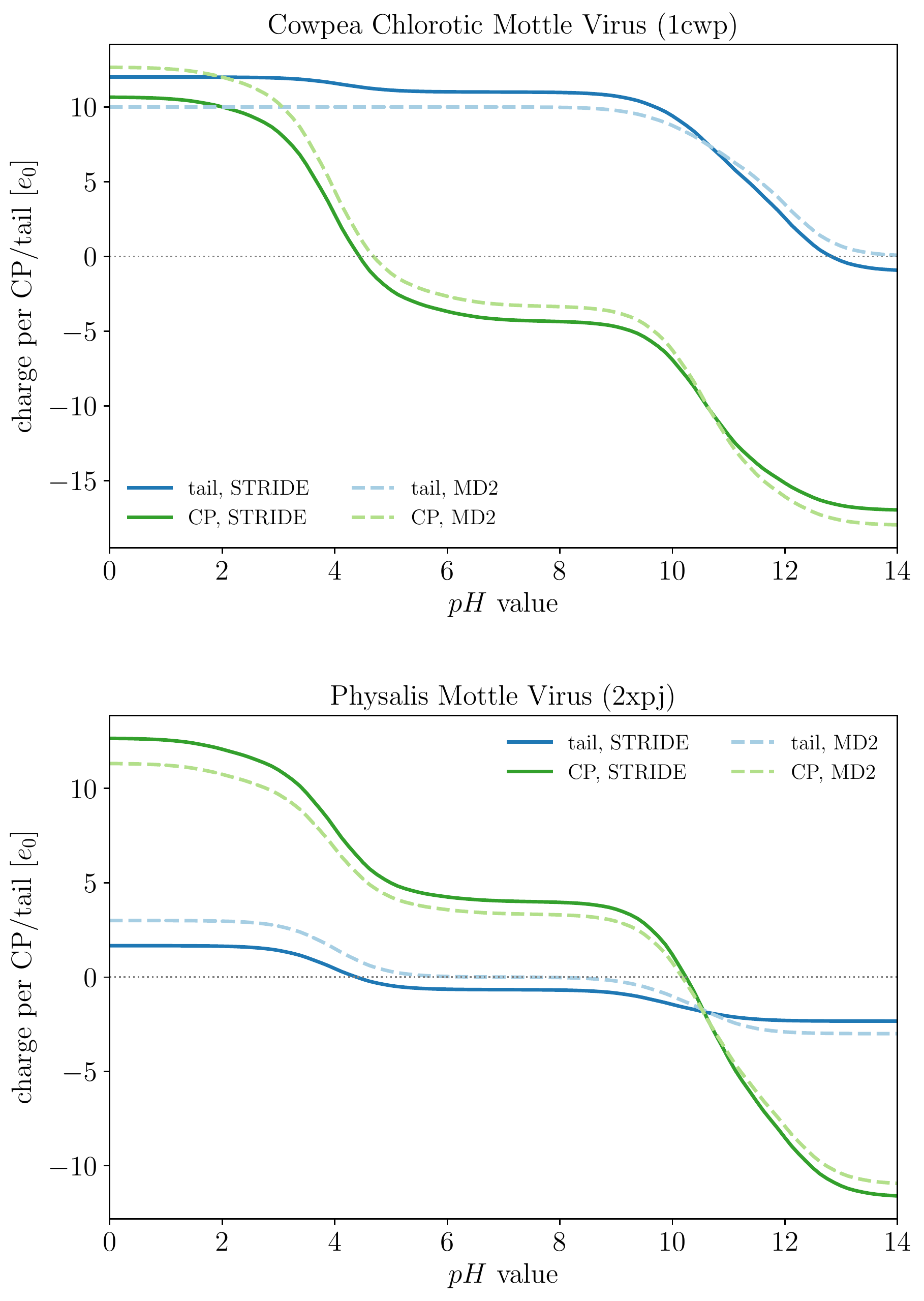}
\end{center}
\caption{$pH$ dependence of the average charge on the N-tails and CPs of CCMV (PDB: 1CWP) and PhMV (PDB: 2XPJ). The dependence is shown for both definitions of the N-tails: using the STRIDE assignment of protein secondary structure or the MD2 prediction of disorder in proteins.
\label{fig:2}}
\end{figure}

The interplay of charge on the N-tails -- especially when they are enriched for positive charge -- and charge on the CPs can be of significant importance for capsid assembly. In particular, certain ssRNA+ viruses tend to preferentially utilize CP-RNA interactions in their assembly, while the capsids of others are stabilized by CP-CP interactions. In Fig.~\ref{fig:2}, we show the $pH$ dependence of charge on the N-tails and CPs of two different viruses, CCMV and Physalis Mottle Virus (PhMV), belonging to Bromoviridae and Tymoviridae families, respectively. In Bromoviridae, positive clusters of charge on the N-tails are known to play an important role in the assembly of functional virions, and the first $26$ residues of CCMV carry a significant positive charge interacting strongly with the negatively charged RNA~\cite{Ni2012,Garmann2014}. Members of Tymoviridae are, on the other hand, predominantly stabilized by CP-CP interactions~\cite{Rao2006}.

The CPs of the two viruses in Fig.~\ref{fig:2} show a similar $pH$ dependence of their charge, which goes from positive to negative as $pH$ is increased from acidic to basic, exhibiting a plateau around neutral $pH$ values -- similar to the dependence previously observed for the charge on full capsids of Leviviridae phages~\cite{Nap2014}. The CP of CCMV has an acidic isoelectric point (point of vanishing charge, where CPs can be crudely treated as electric dipoles~\cite{Bruinsma2016,ALB2017}), whereas the CP of PhMV has a basic one. The charge on the N-tails and its $pH$ dependence are, on the other hand, quite different for the two viruses. The N-tails of CCMV have a large positive charge, mostly stemming from basic AAs and in particular from pronounced ARMs~\cite{Rao2006,Perlmutter2013}, and the large positive charge persists far into the range of basic $pH$ values. Charge on the N-tails of PhMV is comparatively much lower, and becomes negative early on in the $pH$ range. This is true regardless of the definition of the N-tails we use, be it by virtue of secondary structure assignment or protein disorder prediction.

\subsection{Comparing definitions of N-tails: secondary structure and protein disorder}

\begin{figure}[!t]
\begin{center}
\includegraphics[width=\columnwidth]{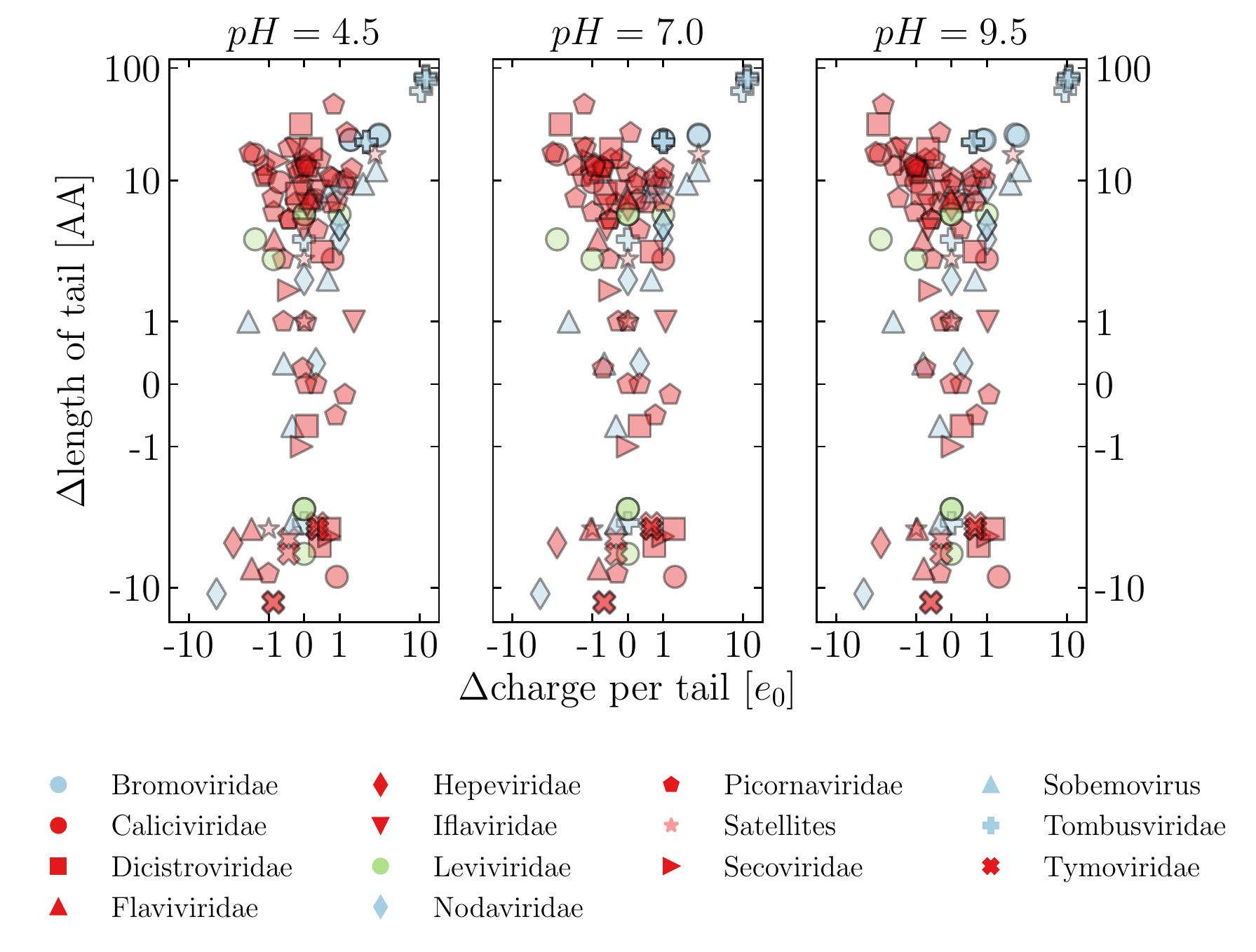}
\end{center}
\caption{Difference in the predictions of the length and the average charge on the N-tails as given by the STRIDE assignment of secondary structure and the MD2 prediction of protein disorder. The charge on the tails is calculated at three different values of $pH$. Blue symbols indicate viral families where N-tails are enriched for clusters of positive charge, and red symbols indicate those where this is not the case (see also the Methods section). In addition, we separately indicate the family Leviviridae (green symbols), the members of which are bacteriophages, and satellite viruses (light red symbols). The same color scheme for the families is used also in other figures in the paper.
\label{fig:3}}
\end{figure}

Before we analyze any further the $pH$ dependence of N-tail and CP charges and their relations to other properties of viruses, we would like to compare more in detail the predictions of the two different definitions of N-tails proposed in the Methods section. These define the tails as either based on the assigned protein secondary structure (STRIDE) or by prediction of intrinsically disordered regions in it (MD2). Figure~\ref{fig:3} shows the differences between the two methods in the predicted tail lengths and the average charge per tail, evaluated at three different $pH$ values, for the entire dataset of analyzed viruses; individual plots of the differences in the predictions for two viruses, BMV and PhMV, are shown in Fig.~\ref{fig:S1} in the Supplementary Material. The example of BMV can also be seen in the sketch of Fig.~\ref{fig:1}, where the tail determined by the first definition is shown in the 3D structure of the CP, showing that it indeed ends at the bulk, structured part of the CP. In addition, the partial AA sequence of the protein already indicates that the major part of positive charge on the tail stems from arginine residues. The tail determined by the second definition is, on the other hand, shorter by $25$ AA, but nonetheless captures the majority of the positively charged clusters on the tail (Fig.~\ref{fig:S1}).

From Fig.~\ref{fig:3} we can see that, for most viruses in our dataset, the differences between the two N-tail definitions in the predicted length of the N-tails are below $20$ AA. In general, tails defined according to the secondary structure assignment tend to be longer compared to the tails from the disorder-based definition. While these differences in tail lengths might still be large enough to change the number of charges on them, it turns out that the differences in the predicted charge on the tails are usually smaller than $\pm2$ $e_0$, and do not seem to depend on the differences in the predicted tail lengths.

\begin{figure*}[!htb]
\begin{center}
\includegraphics[width=0.65\textwidth]{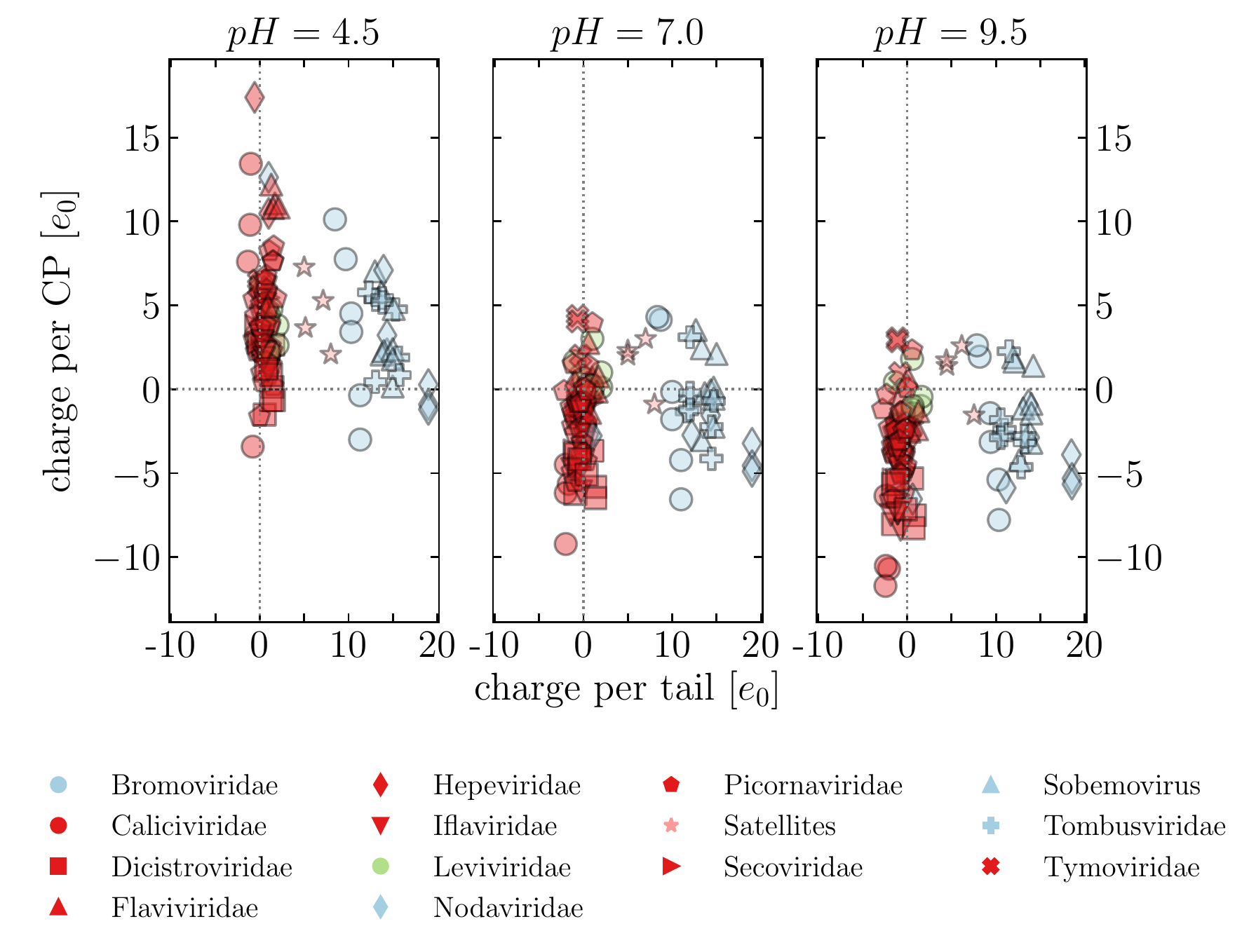}
\end{center}
\caption{Average charge on the N-tails and CPs of viruses in our dataset, evaluated at three different values of $pH=4.5$, $7$, $9.5$. The tails are obtained using the STRIDE assignment of protein secondary structure.
\label{fig:4}}
\end{figure*}

There are, however, three notable exceptions to these observations, all of them belonging to Tombusviridae: Tomato Bushy Stunt Virus (PDB: 2TBV), Melon Necrotic Spot Virus (PDB: 2ZAH), and Cucumber Necrosis Virus (PDB: 4LLF). In all three cases, predictions based on protein disorder result in significantly smaller tails (a difference of more than $80$ AA), which, as a consequence, results in an underprediction of the charge these tails carry, to an extent of almost $10$ $e_0$. Interestingly, the other Tombusviridae entries in our dataset do not show these differences.

In the rest of the paper, we will use the results obtained using the first definition of N-tails (based on secondary structure assignment), keeping in mind that the results obtained using the second definition (based on protein disorder prediction) mostly match those of the first one, yet remembering that there are at the same time a few notable exceptions. For completeness, all the results obtained using the first definition and analyzed in the paper are also shown using the second definition in Figs.~\ref{fig:S2}-\ref{fig:S7} in the Supplementary Material.

\subsection{Charges on N-tails of ssRNA+ viruses}

\begin{figure}[!htb]
\begin{center}
\includegraphics[width=\columnwidth]{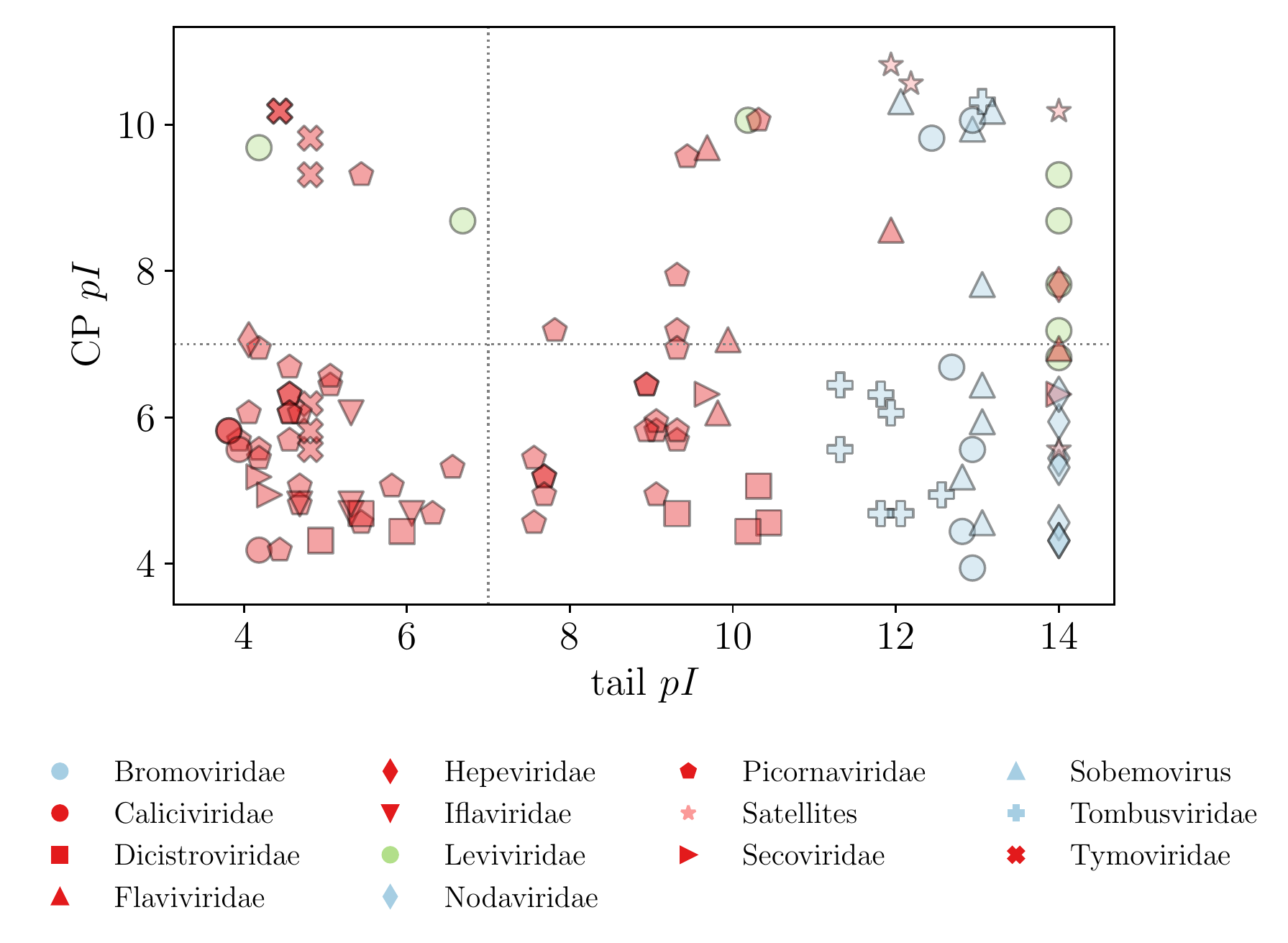}
\end{center}
\caption{Isoelectric points of N-tails and CPs of viruses in our dataset. The tails are obtained using the STRIDE assignment of protein secondary structure.
\label{fig:5}}
\end{figure}

In order to compare the $pH$ dependence of the charge on the N-tails and CPs of all viruses in our dataset, we show in Fig.~\ref{fig:4} the average charge on them at three different values of $pH=4.5$, $7$, $9.5$, and in Fig.~\ref{fig:5} their isoelectric points (cf.\ Fig.~\ref{fig:2}, where the full $pH$ dependence of tail and CP charge is shown for two individual viruses). The plots in Figs.~\ref{fig:4} and~\ref{fig:5} are obtained using the first definition of N-tails (STRIDE); similar plots obtained using the second definition (MD2) are shown in Figs.~\ref{fig:S2} and~\ref{fig:S3}.

\begin{figure*}[!htb]
\begin{center}
\includegraphics[width=0.65\textwidth]{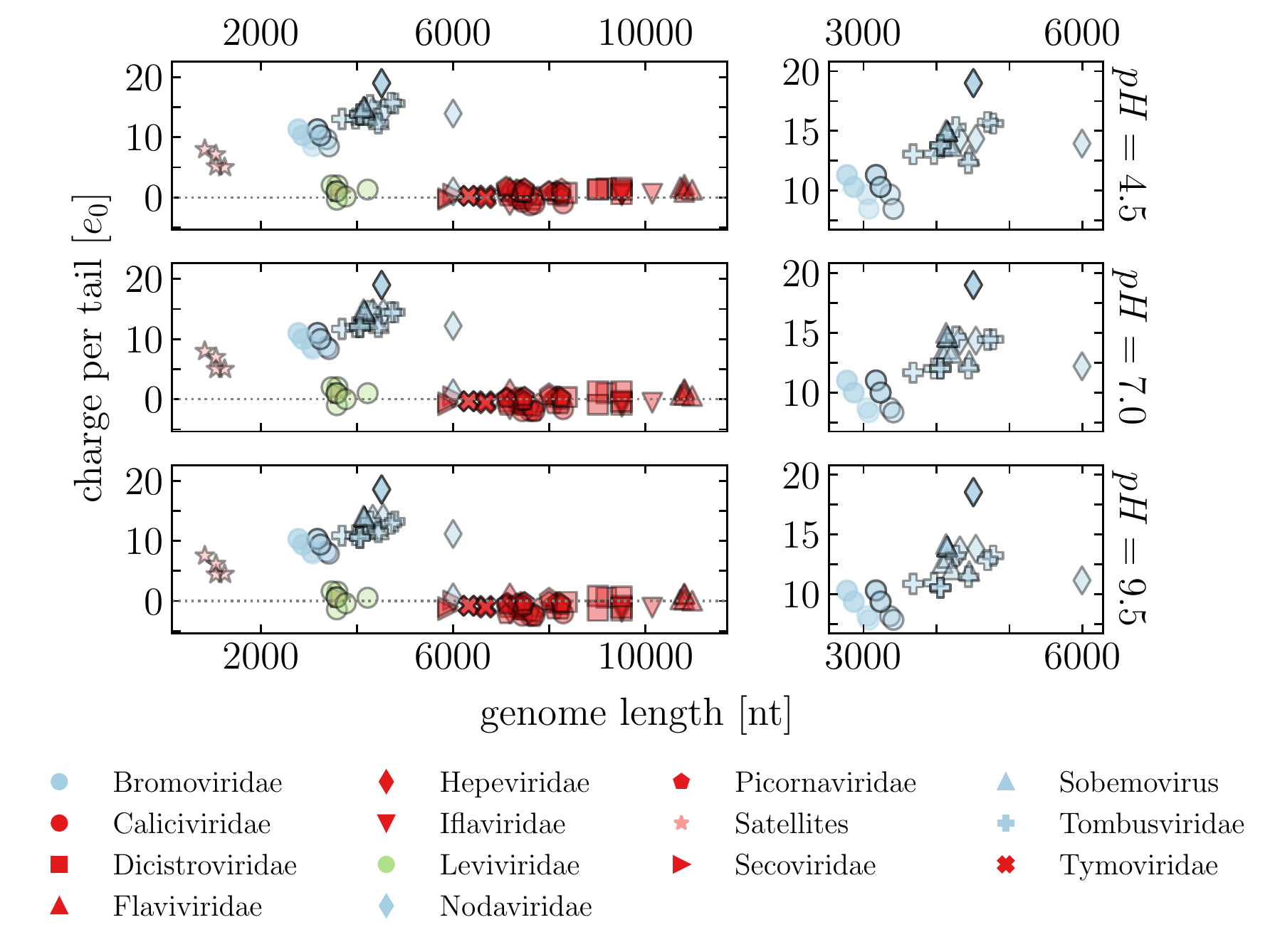}
\end{center}
\caption{Average charge on the N-tails of viruses in our dataset compared to the characteristic lengths of their genomes. The charge is calculated at three different values of $pH$, and the tails are obtained using the STRIDE assignment of protein secondary structure. Smaller plots -- note the different scales -- enlarge the part of the plot where the viral families which utilize positively charged tails are located. Viruses in Bromoviridae family have a tripartite genome and pack the larger two segments, the longer RNA1 and the shorter RNA2, into separate virions. These entries are shown twice, with the shorter genome having a symbol with an edge of a lighter color. Viruses in Nodaviridae family have a bipartite genome, where both RNAs are packaged into a single virion. Genome lengths of these entries show the combined length of both RNAs.
\label{fig:6}}
\end{figure*}

When we plot the average charge on the CPs and N-tails at three different values of $pH$ -- acidic, neutral, and basic -- we see a very clear separation of viral families based on the presence of clusters of positive charge on their tails (Fig.~\ref{fig:4}). While the charge on viral CPs varies in a similar manner across all families, most of the viruses carry close to zero charge on their N-tails. Viruses in some of the families have, however, significantly more positively charged tails: in our dataset, these are the families Bromoviridae, Nodaviridae, Sobemovirus, and Tombusviridae. And while the charge on the CPs shifts overall from positive to negative when the $pH$ increases from acidic to basic, the charge on the N-tails of viruses in these families remains positive and decreases only very slightly. It takes very basic values of $pH$ to eliminate the positive charge these tails carry.

This is signified also by the highly basic $pI$s of the N-tails in these viruses where positive charge on the N-tails plays a significant role ($pI\gtrsim11$ using the first definition of a tail; Fig.~\ref{fig:5}). In addition, this illustrates the fact that these tails consist predominantly of clusters of positive charge only (often as a part of the ARMs~\cite{Perlmutter2015,Rao2006,Schneemann2006}), with negatively charged residues few and far between. On the contrary, $pI$s of the tails of other viruses are in general more acidic and span a larger range of values. We also note that $pI$s of viruses where the N-tails are very short or carry almost no charge fall on the line $pI_{\mathrm{tail}}=14$ as a consequence of the flat $pH$ dependence of their charge. The $pI$s of the CPs also span a large range of values, no matter which family the virus belongs to. The majority is, however, acidic, which then also determines the $pI$ of the entire capsid protein (CP and N-tail together) when the charge on the tail is small. The $pI$s of the entire capsid proteins of viruses where the N-tails carry large positive charge are, in contrast, a combination of the CP and the tail $pI$s. 

\subsection{N-tail charge, genome length, and capsid size}

Lastly, we wish to examine the relationship between the predicted length and charge on the N-tails with some other characteristics of the viruses in our dataset. Specifically, we will be interested in the variable lengths of the viral RNA genomes -- thus bearing a variable net negative charge -- and in the average capsid sizes, which are in icosahedral ssRNA+ viruses cases often tightly related to a characteristic compactness of their genomes~\cite{Yoffe2008,Tubiana2015}.

Figure~\ref{fig:6} compares the average charge on the viral N-tails and the characteristic lengths of their wild-type (WT) genomes. Interestingly and notably, we observe that genome lengths of viruses which utilize positively charged N-tails appear to reach only a limited value ($\sim6$ knt). Other viruses, which do not possess positively charged N-tails, tend to have longer genomes, all the way up to $10$ knt. The only exception of a virus with positively charged tails and a comparable genome length is the recently discovered Orsay virus (PDB: 4NWV), which infects nematodes and bears semblance to viruses in Nodaviridae family, but has not yet been classified~\cite{Guo2014,Franz2014}. The exceptions in the opposite sense are the phages belonging to Leviviridae family, which do not possess any tails and yet pack genomes of only $3$-$4$ knt in length. Some of these phages, such as MS2, are known to utilize RNA packaging signals to direct their capsid assembly~\cite{Twarock2013}. We note that viruses from Togaviridae family, absent in our dataset, are also known to possess N-tails containing clusters of positive charge~\cite{Ni2013,Ni2012,Belyi2006}. While these viruses tend to pack longer genomes ($\sim10$ knt), they also have significantly larger capsids with $\mathcal{T}=4$ symmetry, unlike any virus in our dataset.

In addition, we enlarge in Fig.~\ref{fig:6} the region where the viruses utilizing positively charged tails are located. While some of the viruses here show a distinct correlation between the charge on the viral N-tails and the characteristic lengths of their WT genomes, others appear without any correlation. This is in stark contrast to the claims of universality of the genome-to-tail charge ratio based on theoretical modeling (see Discussion for details).

\begin{figure}[!b]
\begin{center}
\includegraphics[width=\columnwidth]{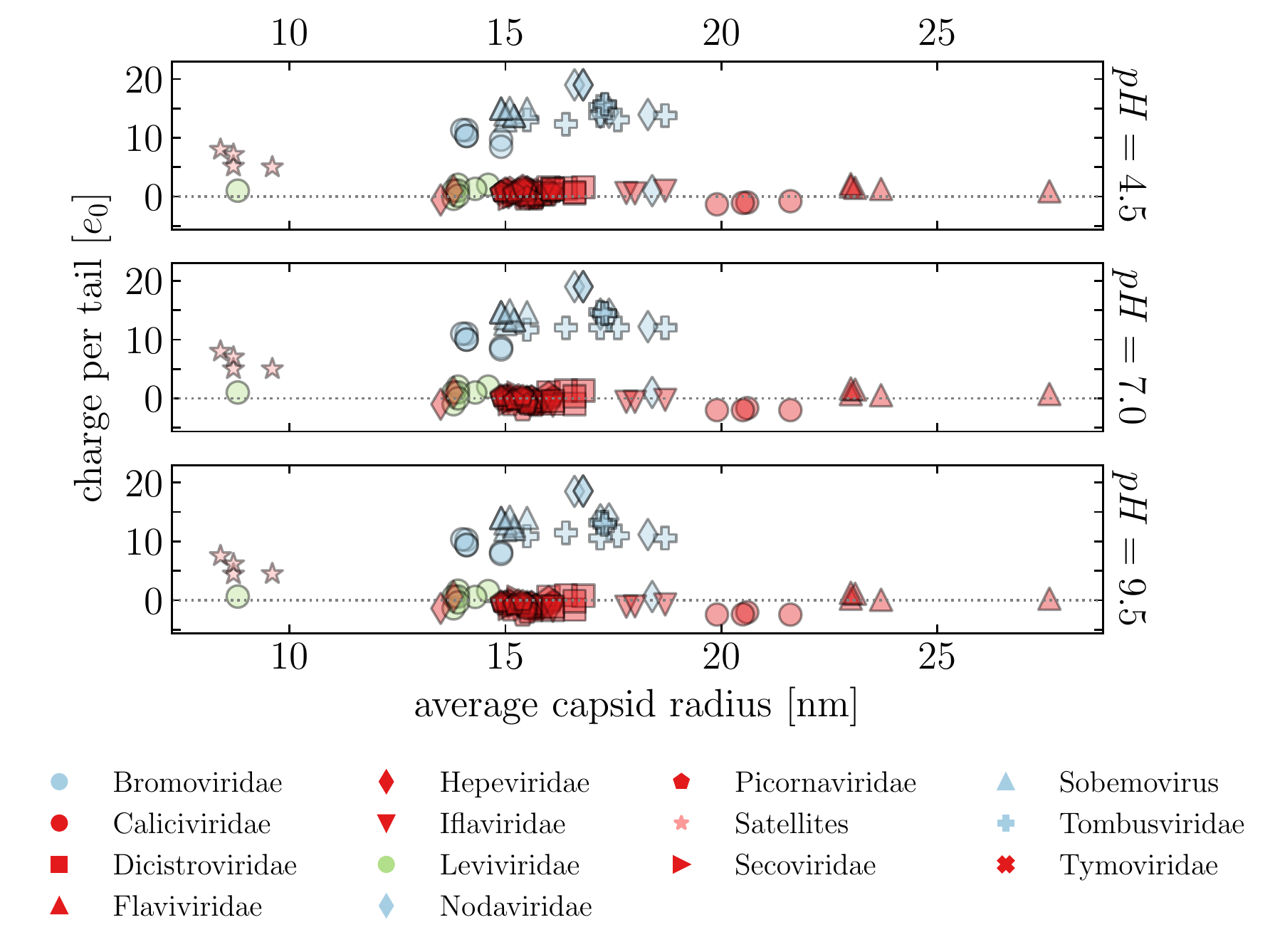}
\end{center}
\caption{Average charge on the N-tails of viruses, calculated at three different values of $pH$, compared to the average radii of their capsids (obtained from VIPERdb). The tails are obtained using the STRIDE assignment of protein secondary structure.
\label{fig:7}}
\end{figure}

In the viruses in our dataset, the more striking relation is thus the one between the genome lengths of viruses which utilize positively charged N-tails and the genome lengths of those which do not, as the latter tend to pack much larger genomes than the former. This large difference is, interestingly, not related to the average size of the capsids into which these genomes are packaged (Fig.~\ref{fig:7}). Vast majority of the capsids have similar average size, even though their genome sizes vary significantly (as does their CP composition~\cite{Rao2006}). The observed difference in the genome sizes of the two broad classes of viruses is also not a consequence of the viruses without any charge on the tails actually having no tails at all -- the tails of the viruses in our dataset range anywhere from $0$-$100$ AA, and the lengths of non-charged tails can still approach $50$ AA (Fig.~\ref{fig:S8} in the Supplementary Material). Similarly, the predicted lengths of the tails also do not correlate with the genome lengths (Fig.~\ref{fig:S9} in the Supplementary Material). Of course, the absence of net charge on the N-tails of viruses with longer genomes does not mean they cannot bind the genome. In these viruses, a more complex electrostatic mechanism could be at work, involving, for instance, polyampholyte-polyelectrolyte complexation~\cite{Dobrynin2008,Cao2015,Jeon2005} or multipolar interactions~\cite{ALB2017,Li2017}.

\section{Discussion}

For many viruses, the non-specific electrostatic interactions between the positively charged, highly flexible N-terminal arms and the RNA genome play a fundamental role in their assembly and structural integrity. However, since these extended CP tail groups lack any definite structure and are thus intrinsically disordered, their definition cannot be entirely devoid of ambiguity. In an attempt to minimize this underlying vagueness, we investigated the length and the charge state of the N-tails of $80$ distinct (and $116$ in total) viruses in detail. We did this by introducing two {\em concise but different definitions} of a capsid protein N-tail, based either on the first occurrence of a given secondary structure element, or on the detection of an intrinsically disordered, contiguous part of the protein at the N-terminus end. In this choice we have attempted to generalize the previous work of Hu {\em et al.}~\cite{Hu2008}, based on a dataset of $27$ viruses, where the N-tail was defined as the flexible sequence of AAs starting from the protein N-terminus and ending at the first $\alpha$-helix (H) or $\beta$-sheet (E). Their work moreover established that the disordered (free) part of the N-tail, obtained from a comparison between the tail sequence and the missing part in the experimental data, entails on the average $76$\% of the full tail length.

The immediate relevance of the length and charge state investigation of the N-tails is most pertinent for the elucidation of the possible {\em universal value of the genome-to-tail charge ratio} in viruses with positively charged tails. In the seminal work of Belyi and Muthukumar~\cite{Belyi2006}, focused on a subclass of $15$ different WT and $5$ mutant ssRNA viruses that bind their genome by using long and highly basic peptide arms, they found a linear scaling between the genome length and the net charge on the capsid peptide arms. This scaling appeared to be robust with very low uncertainty. The error of the charge on the peptide arms was estimated at just $\pm1$ residue and mainly attributed to sequence variations between virus species and the uncertainty in distinguishing flexible peptide arms from the bulk of capsid protein. However, no detailed definition of the tail was provided, and it is unclear whether and how different definitions would modify the main results. Similar conclusion regarding the universality of the genome-to-tail charge ratio was also reached by Hu {\em et al.}~\cite{Hu2008}, who report a scattered charge inversion ratio with a median value $1.8$ for a subset of $13$ from the $27$ viruses described above. Contrasting the claims of universality, a study by Ting {\em et al.}~\cite{Ting2011}, which used a thermodynamic framework to determine the optimal genome length in electrostatically-driven viral encapsidation, led to an opposite conclusion. Namely, they found no universal genome-to-capsid charge ratio, and that a fitted linear relationship between the genome and capsid charge is quite sensitive to the choice of viruses included in the dataset. Nevertheless, all the viruses from the Refs.~\cite{Belyi2006} and~\cite{Hu2008} were found to be overcharged with respect to the packaged genome, a situation which the authors attribute to ``Donnan potential'' and not specifically to the electrostatic attraction between the RNA and the capsomeres~\cite{Siber2008}. In addition, in all these works the RNA was treated as a linear polyelectrolyte, ignoring its secondary structures, which were very recently shown to affect the virus electrostatics profoundly and fundamentally~\cite{Erdemci2014,Erdemci2016,Li2017}.

Regarding the genome-to-tail charge ratio, the results presented in this work -- based on improved and more detailed definitions of the N-tails and their charge, as well as including a large(r) dataset of viruses -- are  more in line with the observations of Ting {\em et al.}~\cite{Ting2011}, and do not point towards a universal scaling of the ratio in viruses with positively charged tails (Fig.~\ref{fig:6}). They do, however, show that the lengths of the WT genomes of these viruses do not seem to exceed $\sim 6$ knt, unlike in other viruses with similar capsid size, although some caution should be exercised since this could easily be affected by the limited amount of viruses in our dataset. In addition, it is known that viruses with positively charged tails can also pack non-native RNAs and other polymers of different lengths~\cite{Hu2008a,Cadena2012}, as is the case with, e.g., CCMV. CCMV preferentially packages RNA1 of BMV instead of its own native RNA1~\cite{Comas2012}, and is found capable of packaging polyU RNAs, which do not form secondary structures and act essentially as structureless linear polymers~\cite{Beren2017}. The polyU RNAs are packaged more efficiently than WT RNAs of equal length, and while polyU RNAs up to $5$ knt are completely packaged, the resuting virions exhibit smaller-than-WT capsids. Longer polyU RNAs up to $9$ knt are also packaged, but into multiplet capsids. The conclusion is therefore that RNA secondary structure (or its absence) plays an essential role in determining the capsid structure during self-assembly of CCMV-like particles, as was later rationalized in detail from RNA polyelectrolyte models taking into account the secondary structure of non-linear RNAs~\cite{Erdemci2016,Erdemci2014}. Inherently branched RNA secondary structure appears to allow viruses to maximize the amount of encapsidated genome, and makes the assembly more efficient and the virion more stable. Nevertheless, at present there is no good rationalization for why so many viruses show absolutely no correlation between the genome and the charge on the N-tails. One could in principle invoke PSs or, in more general terms, the sequence of the viral RNA that could somehow decouple the genomic charge and the structural capsid charge, leading to the features observed in Fig.~\ref{fig:6}.

Apart from the theory-driven analytical polyelectrolyte models of RNA packaging, simulations of viral assembly show that for a given type of CP and solution conditions there exists an optimal length of RNA for the assembly~\cite{Perlmutter2015}. This is confirmed by analytical calculations in numerous works that have studied the various influences on the optimal length of the packaged RNA~\cite{Belyi2006, Hu2008, Siber2008, Ting2011, Erdemci2014, Erdemci2016}. While the analysis of generic electrostatics for a linear RNA in a charged shell is straightforward, the details of the non-uniform charge distribution of the shell and indeed of the CP-tail effect, coupled to the secondary structure of the RNA genome, are more difficult to evaluate quantitatively but appear to underpredict the optimal genome length and thus the overall overcharging of the virion~\cite{Li2017}. In addition, calculations based on linear polyelectrolytes, rather than base-paired nucleic acids, also underpredict the optimal length of the packaged genome, additionally demonstrating the importance of the nucleic acid structure for the assembly~\cite{Perlmutter2013}. There is therefore a growing evidence that the sequence of the viral RNA plays an important role in packaging -- most probably through the coupling between the secondary structure of the RNA and its modifications when subjected to confinement in the virus shell with a complicated non-uniform charge distribution. The relative importance of RNA-RNA contacts compared to RNA-CP contacts and the detailed roles of specific and non-specific (electrostatic) interactions has yet to be determined conclusively.

The results presented in our study -- and specifically, the lack of an observed universal genome-to-tail charge ratio -- are in part constrained by the limited dataset of viruses, even though it is several times larger than those used in previous works. In addition, the results can be influenced to an extent by different definitions of N-tails, as well as the (necessary) simplifications assumed in the calculation of charge on the AA residues of capsid proteins (cf.\ Ref.~\cite{ALB2017} for a detailed discussion of the latter). Lastly, we wish to mention a different possible source of discrepancies related to the choice of the virus dataset, and that is the structural data itself. Different database entries of CPs of the same virus can lack information both on sequence and structural level, to various degrees. In addition, experimental conditions used in determining the structure of the CP can vary -- including the $pH$ of the surrounding solution, temperature, strain of the virus, and not the least the presence or absence of genomic material in the capsid itself. It is known, for example, that the amount of structural disorder found in capsid proteins varies strikingly not only between but also within viral families~\cite{Xue2014}. 

Variations in the resolved structure of a CP can in turn influence our ability to accurately asses the length and charge on the N-tails. For this reason, we included in our dataset also several different database entries of the same viral CP -- where available -- and investigated the resulting variations in the properties of their N-tails. Figure~\ref{fig:S10} in the Supplementary Material shows the examples of two viruses, human rhinovirus and PhMV, with four different entries each. In the case of the human rhinovirus, we include three different strains, which differ slightly already in the length of the full capsid protein. Sizes of PhMV capsid protein are on the other hand the same in all four cases, yet they, too, possess tails of slightly different lengths. As a consequence, we can observe in both examples some variation in the lengths of the determined tails and in the electrostatic properties of the tails and the CPs. The variation remains much in the same range as when we compared the variation between different definitions of N-tails, and this is the case also for other examples we examined. Some entries can again, however, yield quite different results: the $pI$ of human rhinovirus 2, for instance, is acidic, while those of human rhinoviruses 1A and 16 are basic. Such differences are limited mostly to viruses coming from different strains. Importantly, the results for duplicate entries of viruses with positively charged tails that we have examined do not differ much (cf.\ Tables~\ref{tab:S3a} and~\ref{tab:S4a} in the Supplementary Material).

\section{Conclusion}

Our work presents two clear and complementary definitions of the flexible, disordered N-tails of viral coat proteins: the first based on the assignment of secondary structure to the proteins, the other on the predicted intrinsically disordered regions in them. We have shown that the predictions of the two definitions are comparable and, for the most part, consistent. Predicted lengths of the N-tails usually agree within $20$~AA, and consequent predictions of the average charge on the N-tails are within $\pm2$~$e_0$. These differences are of the same order as the uncertainties stemming from the method used to determine the ionizable AAs and their charge~\cite{ALB2017}. And while using different database entries of the same viral coat protein usually yields comparable results, the number of different deposited capsid structures of ssRNA+ viruses remains small enough that the choice and size of the dataset still has the potential to influence quantitative predictions.

Another important improvement on previous works on N-tails that we have included in our study is taking into account the $pH$-dependent charge on all ionizable AAs, giving us the ability to more accurately determine the charge on the N-tails and CPs at any value of $pH$. This dependence shows how big of a game-changer the positively charged N-tails of some viruses can be, as they often contribute as much charge to the protein as the rest of the (structurally-ordered) CP itself, while pushing the $pI$ of the capsid proteins to the basic range of $pH$. Our observations are in line with other studies showing that CP-CP interactions get weaker with increased $pH$, while CP-RNA interactions remain strong by virtue of positively charged N-tails~\cite{Garmann2015,Garmann2014b,Zlotnick2013}. Here, we have shown that this is a robust consequence of the fact that the $pH$ variation of the N-tail charge is much less pronounced than the corresponding variation for the structural part of the CPs. The understanding of the $pH$-dependence of charge in different viruses should have important consequences for the relative interplay of CP-CP and RNA-CP interactions in them, which can be directly related to their assembly mechanisms as well as stability.  

Comparing the charge on the predicted N-tails of viruses in our dataset and the lengths of the corresponding WT genomes packaged in them, we have observed that viruses which utilize positively charged tails and RNA-CP interactions in their assembly pack smaller genomes than viruses where CP-CP interactions are dominant. This observation is not related to the average size of the capsids these genomes are packaged into. Our data also indicate that there is no ``universal'' genome-to-tail charge ratio in viruses with positively charged tails. The mechanisms behind these observations remain unclear at the moment, but recent discoveries point toward the importance of RNA secondary structure for its packaging and interaction with the capsid proteins.

\ack

ALB and RP acknowledge the financial support from the Slovenian Research Agency (research core funding No. (P1-0055)).\\

\bibliographystyle{iopart-num}
\bibliography{references}

\providecommand{\newblock}{}
\begin{thebibliography}{10}
\expandafter\ifx\csname url\endcsname\relax
  \def\url#1{{\tt #1}}\fi
\expandafter\ifx\csname urlprefix\endcsname\relax\def\urlprefix{URL }\fi
\providecommand{\eprint}[2][]{\url{#2}}

\bibitem{Perlmutter2015}
Perlmutter J~D and Hagan M~F 2015 {\em Annu. Rev. Phys. Chem.\/} {\bf 66}
  217--239

\bibitem{Siber2012}
\v{S}iber A, {Lo\v{s}dorfer Bo\v{z}i\v{c}} A and Podgornik R 2012 {\em Phys.
  Chem. Chem. Phys.\/} {\bf 14} 3746--3765

\bibitem{Sun2010}
Sun S, Rao V~B and Rossmann M~G 2010 {\em Curr. Op. Struct. Biol.\/} {\bf 20}
  114--120

\bibitem{Ni2013}
Ni P and Kao C~C 2013 {\em Virology\/} {\bf 446} 123--132

\bibitem{Rao2006}
Rao A 2006 {\em Annu. Rev. Phytopathol.\/} {\bf 44} 61--87

\bibitem{Siber2008}
{\v{S}}iber A and Podgornik R 2008 {\em Phys. Rev. E\/} {\bf 78} 051915

\bibitem{Schneemann2006}
Schneemann A 2006 {\em Annu. Rev. Microbiol.\/} {\bf 60} 51--67

\bibitem{Beren2017}
Beren C, Dreesens L~L, Liu K~N, Knobler C~M and Gelbart W~M 2017 {\em Biophys.
  J.\/} {\bf 113} 339--347

\bibitem{Erdemci2016}
Erdemci-Tandogan G, Wagner J, van~der Schoot P, Podgornik R and Zandi R 2016
  {\em Phys. Rev. E\/} {\bf 94} 022408

\bibitem{Erdemci2014}
Erdemci-Tandogan G, Wagner J, Van Der~Schoot P, Podgornik R and Zandi R 2014
  {\em Phys. Rev. E\/} {\bf 89} 032707

\bibitem{Yoffe2008}
Yoffe A~M, Prinsen P, Gopal A, Knobler C~M, Gelbart W~M and {Ben-Shaul} A 2008
  {\em Proc. Natl. Acad. Sci. USA\/} {\bf 105} 16153

\bibitem{Tubiana2015}
Tubiana L, {Lo\v{s}dorfer Bo\v{z}i\v{c}} A, Micheletti C and Podgornik R 2015
  {\em Biophys. J.\/} {\bf 108} 194--202

\bibitem{Garmann2014}
Garmann R~F, Comas-Garcia M, Koay M~S, Cornelissen J~J, Knobler C~M and Gelbart
  W~M 2014 {\em J. Virol.\/} {\bf 88} 10472--10479

\bibitem{Belyi2006}
Belyi V~A and Muthukumar M 2006 {\em Proc. Natl. Acad. Sci. USA\/} {\bf 103}
  17174--17178

\bibitem{Hu2008}
Hu T, Zhang R and Shklovskii B 2008 {\em Physica A\/} {\bf 387} 3059--3064

\bibitem{Ting2011}
Ting C~L, Wu J and Wang Z~G 2011 {\em Proc. Natl. Acad. Sci. USA\/} {\bf 108}
  16986--16991

\bibitem{Bruinsma2016}
Bruinsma R~F, Comas-Garcia M, Garmann R~F and Grosberg A~Y 2016 {\em Phys. Rev.
  E\/} {\bf 93} 032405

\bibitem{Ni2012}
Ni P, Wang Z, Ma X, Das N~C, Sokol P, Chiu W, Dragnea B, Hagan M and Kao C~C
  2012 {\em J. Mol. Biol.\/} {\bf 419} 284--300

\bibitem{Xue2014}
Xue B, Blocquel D, Habchi J, Uversky A~V, Kurgan L, Uversky V~N and Longhi S
  2014 {\em Chem. Rev.\/} {\bf 114} 6880--6911

\bibitem{Ford2013}
Ford R~J, Barker A~M, Bakker S~E, Coutts R~H, Ranson N~A, Phillips S~E, Pearson
  A~R and Stockley P~G 2013 {\em J. Mol. Biol.\/} {\bf 425} 1050--1064

\bibitem{Tokuriki2009}
Tokuriki N, Oldfield C~J, Uversky V~N, Berezovsky I~N and Tawfik D~S 2009 {\em
  Trends Biochem. Sci.\/} {\bf 34} 53--59

\bibitem{Perlmutter2015b}
Perlmutter J~D and Hagan M~F 2015 {\em J. Mol. Biol.\/} {\bf 427} 2451--2467

\bibitem{Perlmutter2013}
Perlmutter J~D, Qiao C and Hagan M~F 2013 {\em eLife\/} {\bf 2} e00632

\bibitem{Lavelle2009}
Lavelle L, Gingery M, Phillips M, Gelbart W, Knobler C, Cadena-Nava R,
  Vega-Acosta J, Pinedo-Torres L and Ruiz-Garcia J 2009 {\em J. Phys. Chem.
  B\/} {\bf 113} 3813--3819

\bibitem{Comas2014}
Comas-Garcia M, Garmann R~F, Singaram S~W, Ben-Shaul A, Knobler C~M and Gelbart
  W~M 2014 {\em J. Phys. Chem. B.\/} {\bf 118} 7510--7519

\bibitem{Garmann2015}
Garmann R~F, Comas-Garcia M, Knobler C~M and Gelbart W~M 2015 {\em Acc. Chem.
  Res.\/} {\bf 49} 48--55

\bibitem{Wilts2015}
Wilts B~D, Schaap I~A and Schmidt C~F 2015 {\em Biophys. J.\/} {\bf 108}
  2541--2549

\bibitem{PDB}
Berman H~M, Westbrook J, Feng Z, Gilliland G, Bhat T~N, Weissig H, Shindyalov
  I~N and Bourne P~E 2000 {\em Nucleic Acids Res.\/} {\bf 28} 235--242

\bibitem{VIPERdb}
Carrillo-Tripp M, Shepherd C~M, Borelli I~A, Venkataraman S, Lander G,
  Natarajan P, Johnson J~E, Brooks C~L and Reddy V~S 2009 {\em Nucleic Acids
  Res.\/} {\bf 37} D436--D442

\bibitem{NCBI}
{National Centre for Biotechnology Information (NCBI)} 1988 {Nucleotide
  Database} {Accessed: August 2017}
  \urlprefix\url{https://www.ncbi.nlm.nih.gov/nucleotide/}

\bibitem{UNIPROT}
{UniProt Consortium} {\em et~al.\/} 2017 {\em Nucleic Acids Res.\/} {\bf 45}
  D158--D169

\bibitem{VIRALZONE}
Hulo C, De~Castro E, Masson P, Bougueleret L, Bairoch A, Xenarios I and
  Le~Mercier P 2010 {\em Nucleic Acids Res.\/} {\bf 39} D576--D582

\bibitem{Marshall2001}
Marshall D and Schneemann A 2001 {\em Virology\/} {\bf 285} 165--175

\bibitem{Odegard2010}
Odegard A, Banerjee M and Johnson J~E 2010 {Flock House Virus}: a model system
  for understanding non-enveloped virus entry and membrane penetration {\em
  Cell Entry by Non-Enveloped Viruses\/} ed Johnson J~E (Springer) pp 1--22

\bibitem{STRIDE}
Frishman D and Argos P 1995 {\em Proteins: Struct., Funct., Bioinf.\/} {\bf 23}
  566--579

\bibitem{DSSP}
Kabsch W and Sander C 1983 {\em Biopolymers\/} {\bf 22} 2577--2637

\bibitem{Fodje2002}
Fodje M and Al-Karadaghi S 2002 {\em Protein Eng.\/} {\bf 15} 353--358

\bibitem{Kozlowski2012}
Kozlowski L~P and Bujnicki J~M 2012 {\em BMC Bioinformatics\/} {\bf 13} 111

\bibitem{ALB2017}
{Lo\v{s}dorfer Bo\v{z}i\v{c}} A and Podgornik R 2017 {\em Biophys. J.\/} {\bf
  113} 1454--1465

\bibitem{Nap2014}
Nap R~J, {Lo\v{s}dorfer Bo\v{z}i\v{c}} A, Szleifer I and Podgornik R 2014 {\em
  Biophys. J.\/} {\bf 107} 1970--1979

\bibitem{Guo2014}
Guo Y~R, Hryc C~F, Jakana J, Jiang H, Wang D, Chiu W, Zhong W and Tao Y~J 2014
  {\em Proc. Natl. Acad. Sci. USA\/} {\bf 111} 12781--12786

\bibitem{Franz2014}
Franz C~J, Renshaw H, Frezal L, Jiang Y, F{\'e}lix M~A and Wang D 2014 {\em
  Virology\/} {\bf 448} 255--264

\bibitem{Twarock2013}
Stockley P~G, Twarock R, Bakker S~E, Barker A~M, Borodavka A, Dykeman E, Ford
  R~J, Pearson A~R, Phillips S~E~V, Ranson N~A and Tuma R 2013 {\em J. Biol.
  Phys.\/} {\bf 39} 277–287

\bibitem{Dobrynin2008}
Dobrynin A~V 2008 {\em Curr. Op. Colloid Interface Sci.\/} {\bf 13} 376--388

\bibitem{Cao2015}
Cao Q and You H 2015 {\em Langmuir\/} {\bf 31} 6375--6384

\bibitem{Jeon2005}
Jeon J and Dobrynin A~V 2005 {\em Macromolecules\/} {\bf 38} 5300--5312

\bibitem{Li2017}
Li S, Erdemci-Tandogan G, Wagner J, van~der Schoot P and Zandi R 2017 {\em
  Phys. Rev. E\/} {\bf 96} 022401

\bibitem{Hu2008a}
Hu Y, Zandi R, Anavitarte A, Knobler C~M and Gelbart W~M 2008 {\em Biophys.
  J.\/} {\bf 94} 1428--1436

\bibitem{Cadena2012}
Cadena-Nava R~D, Comas-Garcia M, Garmann R~F, Rao A, Knobler C~M and Gelbart
  W~M 2012 {\em J. Virol.\/} {\bf 86} 3318--3326

\bibitem{Comas2012}
Comas-Garcia M, Cadena-Nava R~D, Rao A, Knobler C~M and Gelbart W~M 2012 {\em
  J. Virol.\/} {\bf 86} 12271--12282

\bibitem{Garmann2014b}
Garmann R~F, Comas-Garcia M, Gopal A, Knobler C~M and Gelbart W~M 2014 {\em J.
  Mol. Biol.\/} {\bf 426} 1050--1060

\bibitem{Zlotnick2013}
Zlotnick A, Porterfield J~Z and Wang J~C~Y 2013 {\em Biophys. J.\/} {\bf 104}
  1595--1604

\end{thebibliography}

\newpage


\section*{Supplementary material}

# Supplementary Material:
# Varieties of charge distributions in coat proteins of ssRNA+ viruses

Anže Lošdorfer Božič[1, *] and Rudolf Podgornik[1, 2]

[1]*Department of Theoretical Physics, Jožef Stefan Institute, SI-1000 Ljubljana, Slovenia*
[2]*Department of Physics, Faculty of Mathematics and Physics,*
*University of Ljubljana, SI-1000 Ljubljana, Slovenia*
(Dated: November 21, 2017)

---

[*] anze.bozic@ijs.si; Corresponding author

# I. COMPARING DEFINITIONS OF N-TAILS: STRIDE AND MD2

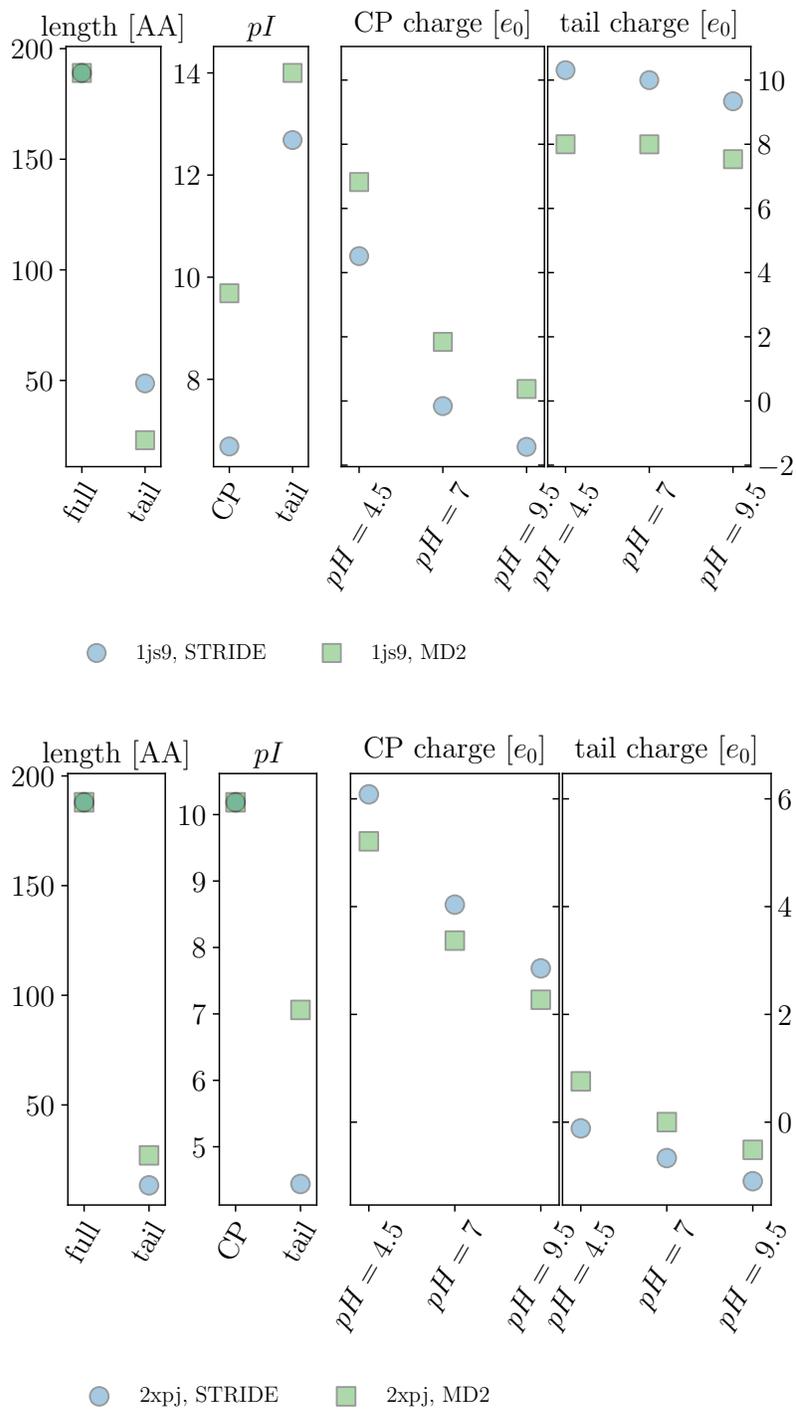

Figure S1. Comparison of the two N-tail definitions used in the main paper (via STRIDE secondary structure assignment and MD2 protein disorder prediction). We show the resulting lengths of the N-tails, the $pI$ of the CPs and the tails, and the average charge on the CPs and tails at three different values of $pH$. Shown for BMV (PDB: 1JS9) and PhMV (PDB: 2XPJ).

## II.  RESULTS OF MD2 PREDICTIONS

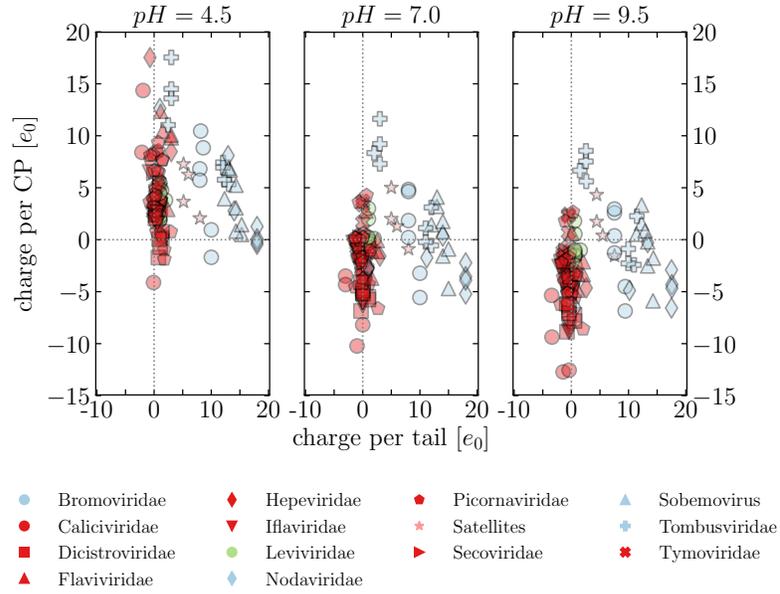

Figure S2.  Average charge on the N-tails and CPs of viruses in our dataset, evaluated at three different values of $pH = 4.5$, 7, 9.5.  The tails are obtained using the MD2 prediction of disorder in proteins.

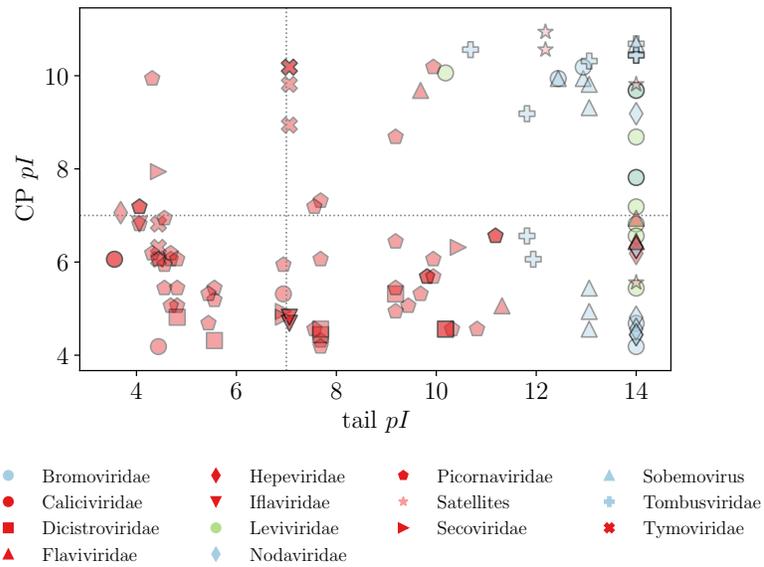

Figure S3.  Isoelectric points of N-tails and CPs of viruses in our dataset.  The tails are obtained using the MD2 prediction of disorder in proteins.

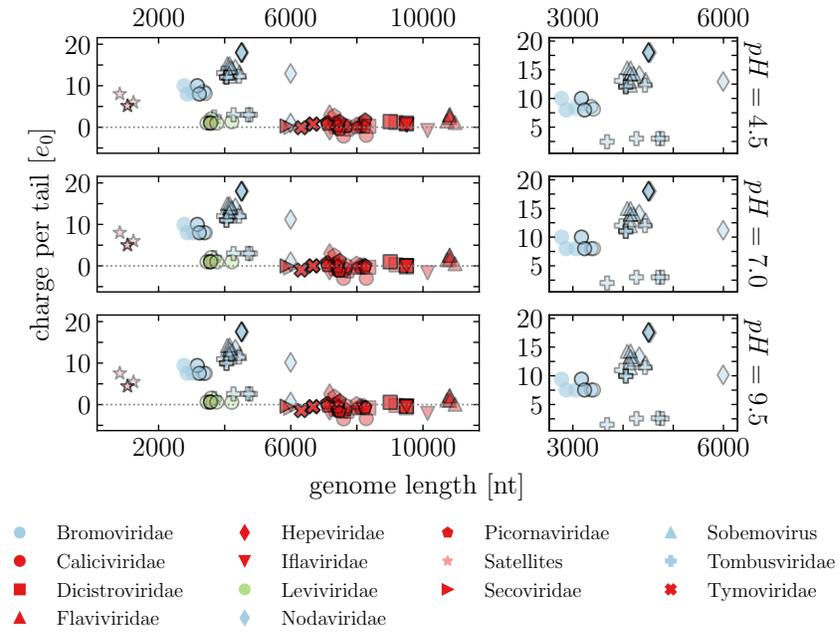

Figure S4. Average charge on the N-tails of viruses in our dataset compared to the characteristic lengths of their genomes. The charge is calculated at three different values of $pH$, and the tails are obtained using the MD2 prediction of disorder in proteins. The smaller plots enlarge the part of the plot where the viral families which utilize positively charged tails are located. Viruses in the Bromoviridae family have a tripartite genome and pack the larger two segments, the longer RNA1 and the shorter RNA2, into separate virions. These entries are shown twice, with the shorter genome having a marker with an edge of a lighter color. Viruses in the Nodoviridae family have a bipartite genome, where both RNAs are packaged into a single virion. The genome length of these entries show the combined length of both RNAs.

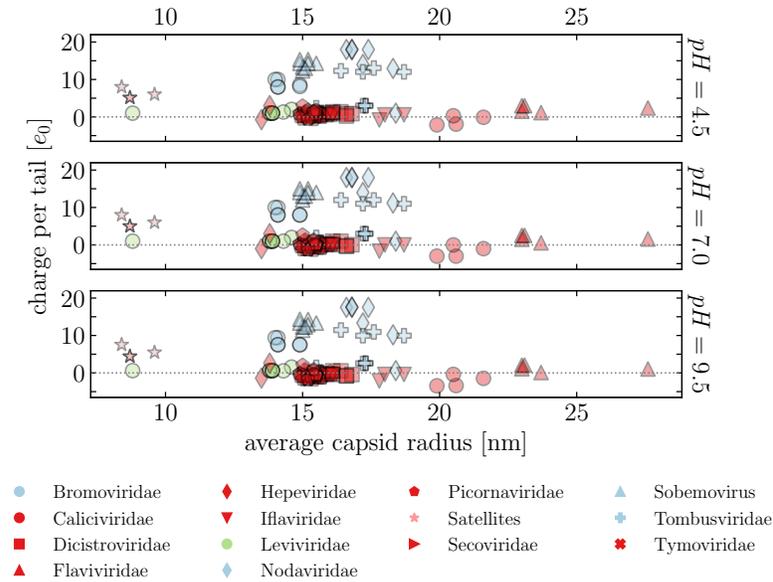

Figure S5. Average charge on the N-tails of viruses, calculated at three different values of $pH$, compared to the average radii of their capsids (obtained from VIPERdb). The tails are obtained using the MD2 prediction of disorder in proteins.

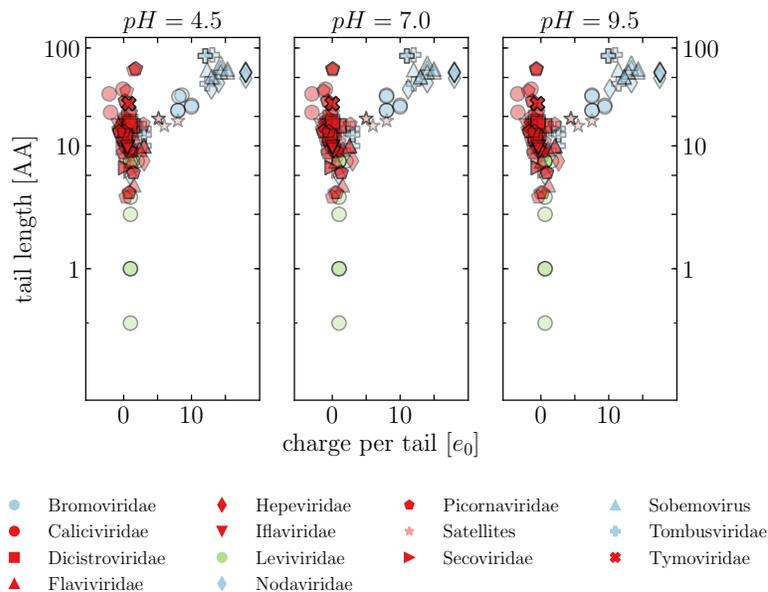

Figure S6. Lengths of the N-tails compared to their average charge. The tails are obtained using the MD2 prediction of disorder in proteins.

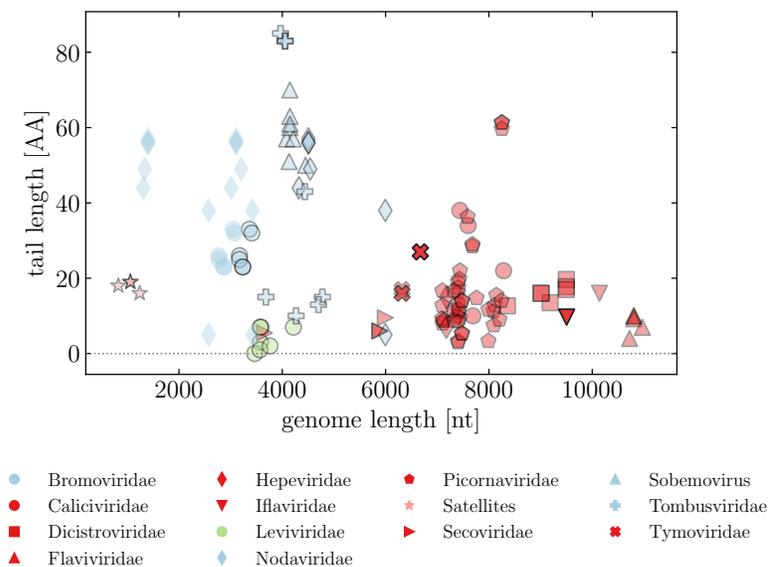

Figure S7. Lengths of the N-tails compared to the lengths of the viral genomes. Viruses in the Bromoviridae family have a tripartite genome and pack the larger two segments, the longer RNA1 and the shorter RNA2, into separate virions. These entries are shown twice, with the shorter genome having a marker with an edge of a lighter color. Viruses in the Nodoviridae family have a bipartite genome, where both RNAs are packaged into a single virion. The genome length of these entries show the combined length of both RNAs. The tails are obtained using the MD2 prediction of disorder in proteins.

## III. LENGTHS OF N-TAILS

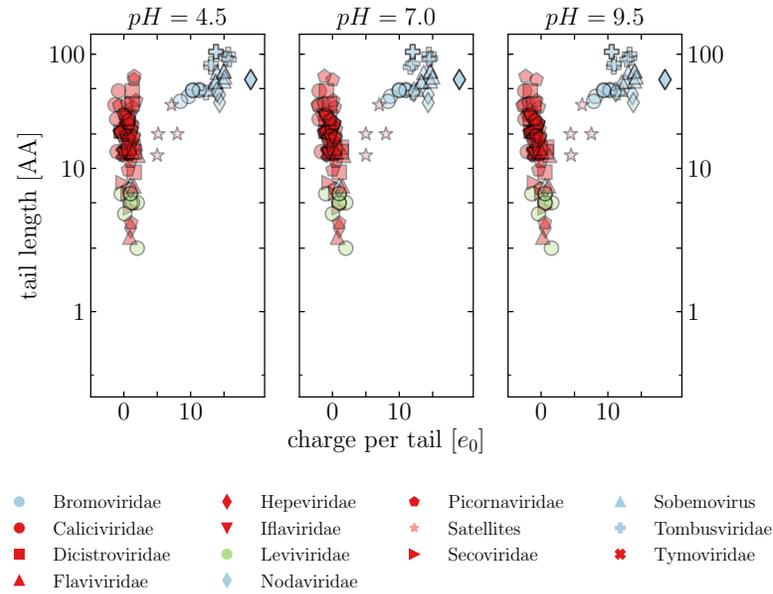

Figure S8. Lengths of the N-tails compared to their average charge. The tails are obtained using the STRIDE assignment of protein secondary structure.

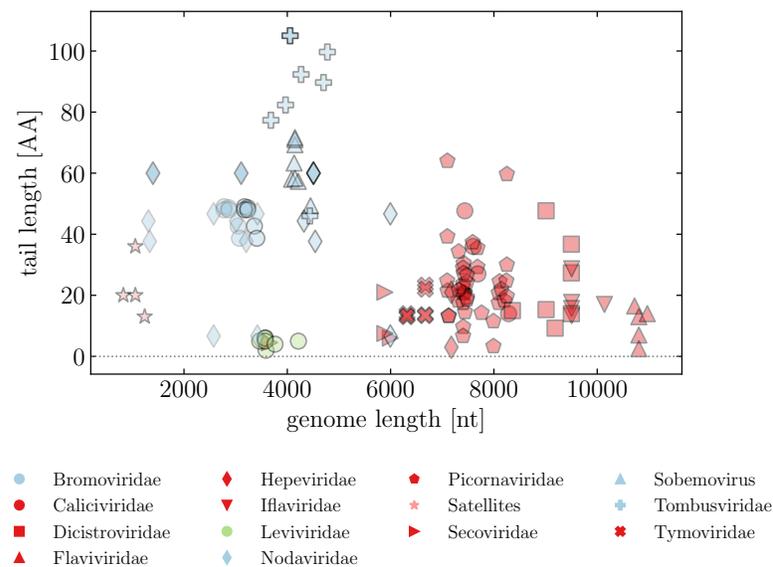

Figure S9. Lengths of the N-tails compared to the lengths of the viral genomes. Viruses in the Bromoviridae family have a tripartite genome and pack the larger two segments, the longer RNA1 and the shorter RNA2, into separate virions. These entries are shown twice, with the shorter genome having a marker with an edge of a lighter color. Viruses in the Nodoviridae family have a bipartite genome, where both RNAs are packaged into a single virion. The genome length of these entries show the combined length of both RNAs. The tails are obtained using the STRIDE assignment of protein secondary structure.

## IV.    COMPARING TAIL PROPERTIES OF SEVERAL DIFFERENT ENTRIES OF THE SAME VIRUS

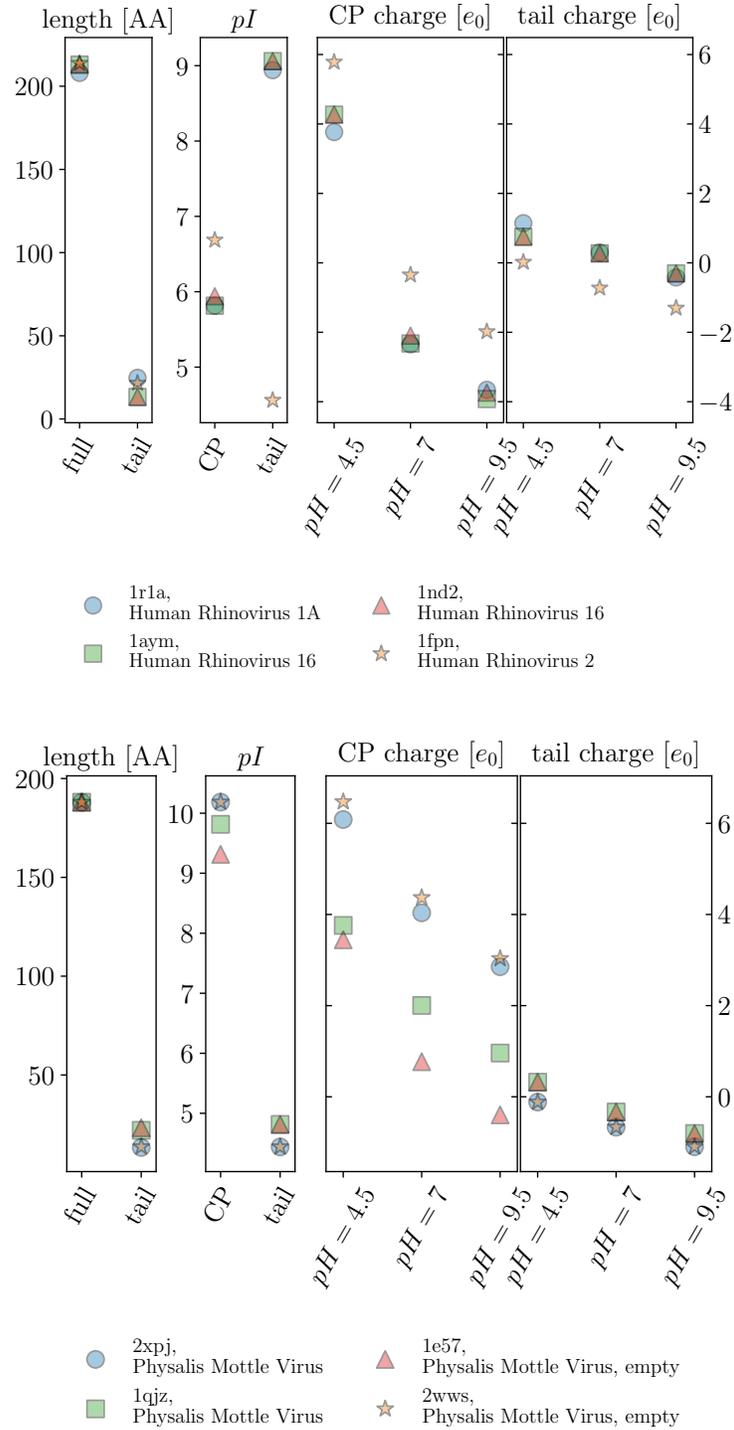

Figure S10.  Comparison of N-tail and CP properties of several different PDB/VIPERdb database entries of the same virus. Shown for human rhinovirus and PhMV. The tails are obtained using the STRIDE assignment of protein secondary structure.

# V.   DATA TABLES

| PDB ID | Virus (entry) | Family | Genus |
|--------|---------------|--------|-------|
| 1js9 | Brome Mosaic Virus | Bromoviridae | Bromovirus |
| 3j7l | Brome Mosaic Virus | Bromoviridae | Bromovirus |
| 1cwp | Cowpea Chlorotic Mottle Virus | Bromoviridae | Bromovirus |
| 1za7 | Cowpea Chlorotic Mottle Virus, mutant | Bromoviridae | Bromovirus |
| 1f15 | Cucumber Mosaic Virus | Bromoviridae | Cucumovirus |
| 1laj | Tomato Aspermy Virus | Bromoviridae | Cucumovirus |
| 4ftb | Flock House Virus | Nodaviridae | Alphanodavirus |
| 4fsj | Flock House VLP | Nodaviridae | Alphanodavirus |
| 1nov | Nodamura Virus | Nodaviridae | Alphanodavirus |
| 2bbv | Black Beetle Virus | Nodaviridae | Alphanodavirus |
| 1f8v | Pariacoto Virus | Nodaviridae | Alphanodavirus |
| 4nwv | Orsay VLP | Nodaviridae | Unclassified |
| 4nww | Orsay VLP, deletion mutant | Nodaviridae | Unclassified |
| 2izw | Ryegrass Mottle Virus | Sobemovirus | Sobemovirus |
| 1smv | Sesbania Mosaic Virus | Sobemovirus | Sobemovirus |
| 2vq0 | Sesbania Mosaic Virus, deletion mutant | Sobemovirus | Sobemovirus |
| 1x35 | Sesbania Mosaic Virus, mutant | Sobemovirus | Sobemovirus |
| 1x33 | Sesbania Mosaic Virus, recombinant | Sobemovirus | Sobemovirus |
| 1f2n | Rice Yellow Mottle Virus | Sobemovirus | Sobemovirus |
| 4sbv | Southern Bean Mosaic Virus | Sobemovirus | Sobemovirus |
| 1ng0 | Cocksfoot Mottle Virus | Sobemovirus | Sobemovirus |
| 1opo | Carnation Mottle Virus | Tombusviridae | Carmovirus |
| 2zah | Melon Necrotic Spot Virus | Tombusviridae | Carmovirus |
| 3jb8 | Maize Chlorotic Mottle Virus | Tombusviridae | Machlomovirus |
| 1c8n | Tobacco Necrosis Virus | Tombusviridae | Necrovirus |
| 4llf | Cucumber Necrosis Virus | Tombusviridae | Tombusvirus |
| 2tbv | Tomato Bushy Stunt Virus | Tombusviridae | Tombusvirus |
| 3zx8 | Turnip Crinkle Virus | Tombusviridae | Carmovirus |
| 3zx9 | Turnip Crinkle Virus, expanded | Tombusviridae | Carmovirus |
| 3j1p | Rabbit Hemmorrhagic Disease Virus | Caliciviridae | Lagovirus |
| 1ihm | Norwalk Virus | Caliciviridae | Norovirus |
| 2gh8 | San Miguel Sea Lion Calicivirus | Caliciviridae | Vesivirus |
| 3m8l | Feline Calicivirus | Caliciviridae | Vesivirus |
| 5cdc | Israel Acute Paralysis Virus | Dicistroviridae | Aparavirus |
| 5lwg | Israel Acute Paralysis Virus, heated – full | Dicistroviridae | Aparavirus |
| 5lwi | Israel Acute Paralysis Virus, heated – empty | Dicistroviridae | Aparavirus |
| 1b35 | Cricket Paralysis Virus | Dicistroviridae | Cripavirus |
| 3nap | Triatoma Virus | Dicistroviridae | Cripavirus |
| 5l7o | Triatoma Virus, empty | Dicistroviridae | Cripavirus |
| 5mqc | Black Queen Cell Virus | Dicistroviridae | Triatovirus |
| 5u4w | Zika Virus, immature | Flaviviridae | Flavivirus |
| 5iz7 | Zika Virus, thermally stable | Flaviviridae | Flavivirus |
| 5ire | Zika Virus | Flaviviridae | Flavivirus |
| 3j27 | Dengue Virus | Flaviviridae | Flavivirus |
| 5wsn | Japanese Encephalitis Virus | Flaviviridae | Flavivirus |
| 2ztn | Hepatitis E VLP, ORF2 | Hepeviridae | Hepevirus |
| 3hag | Hepatitis E VLP | Hepeviridae | Hepevirus |
| 5j98 | Slow Bee Paralysis Virus | Iflaviridae | Iflavirus |
| 5j96 | Slow Bee Paralysis Virus | Iflaviridae | Iflavirus |
| 5lk7 | Slow Bee Paralysis Virus, $pH = 5.5$ | Iflaviridae | Iflavirus |
| 5lk8 | Slow Bee Paralysis Virus, empty | Iflaviridae | Iflavirus |
| 5g52 | Deformed Wing Virus | Iflaviridae | Iflavirus |

Table S1. List of viruses in the dataset. Shown are their PDB [1]/VIPERdb [2] ID, the name of the virus together with any specific characteristics (strain, mutant, ...) as deposited in the database, the viral family, and genus.

| PDB ID | Virus (entry) | Family | Genus |
|--------|---------------|--------|-------|
| 5ned | Foot-and-mouth-disease Virus O | Picornaviridae | Aphthovirus |
| 4gh4 | Foot-and-mouth-disease Virus A22 | Picornaviridae | Aphthovirus |
| 1fmd | Foot-and-mouth-disease Virus C | Picornaviridae | Aphthovirus |
| 2wzr | Foot-and-mouth-disease Virus SAT1 | Picornaviridae | Aphthovirus |
| 2wff | Equine Rhinitis A Virus | Picornaviridae | Aphthovirus |
| 2ws9 | Equine Rhinitis A Virus, low $pH$ | Picornaviridae | Aphthovirus |
| 5cfc | Saffold Virus 3 | Picornaviridae | Cardiovirus |
| 5a8f | Saffold Virus 3 | Picornaviridae | Cardiovirus |
| 2mev | Mengo Virus | Picornaviridae | Cardiovirus |
| 1tme | Theiler Virus, DA strain | Picornaviridae | Cardiovirus |
| 1tmf | Theiler Virus, BeAn strain | Picornaviridae | Cardiovirus |
| 1bev | Bovine Enterovirus | Picornaviridae | Enterovirus |
| 3vbu | Human Enterovirus 71, empty | Picornaviridae | Enterovirus |
| 3vbs | Human Enterovirus 71 | Picornaviridae | Enterovirus |
| 4aed | Human Enterovirus 71 | Picornaviridae | Enterovirus |
| 4wm8 | Human Enterovirus D68 | Picornaviridae | Enterovirus |
| 4biq | Coxsackievirus A7, empty | Picornaviridae | Enterovirus |
| 1cov | Coxsackievirus B3 | Picornaviridae | Enterovirus |
| 5c4w | Coxsackievirus A16 | Picornaviridae | Enterovirus |
| 1z7s | Coxsackievirus A21 | Picornaviridae | Enterovirus |
| 1hxs | Poliovirus, Mahoney strain | Picornaviridae | Enterovirus |
| 5k0u | Human Rhinovirus C | Picornaviridae | Enterovirus |
| 5jzg | Human Rhinovirus C, empty | Picornaviridae | Enterovirus |
| 2x5i | Echovirus 7 | Picornaviridae | Enterovirus |
| 1h8t | Echovirus 11 | Picornaviridae | Enterovirus |
| 1mqt | Swine Vesicular Disease Virus | Picornaviridae | Enterovirus |
| 1oop | Swine Vesicular Disease Virus | Picornaviridae | Enterovirus |
| 5wtf | Hepatitis A Virus, complexed (FAB) | Picornaviridae | Hepatovirus |
| 4qpg | Hepatitis A Virus, empty | Picornaviridae | Hepatovirus |
| 5wte | Hepatitis A Virus | Picornaviridae | Hepatovirus |
| 4qpi | Hepatitis A Virus | Picornaviridae | Hepatovirus |
| 5lvc | Aichi Virus 1, empty | Picornaviridae | Kobuvirus |
| 5gka | Aichi Virus | Picornaviridae | Kobuvirus |
| 5aoo | Aichi Virus A | Picornaviridae | Kobuvirus |
| 4z92 | Human Parechovirus 1 | Picornaviridae | Parechovirus |
| 3jb4 | Ljungan Virus | Picornaviridae | Parechovirus |
| 1r1a | Human Rhinovirus 1A | Picornaviridae | Rhinovirus |
| 1aym | Human Rhinovirus 16 | Picornaviridae | Rhinovirus |
| 1nd2 | Human Rhinovirus 16 | Picornaviridae | Rhinovirus |
| 1fpn | Human Rhinovirus 2 | Picornaviridae | Rhinovirus |
| 3cji | Seneca Valley Virus | Picornaviridae | Senecavirus |
| 1pgw | Bean Pod Mottle Virus, top component | Secoviridae | Comovirus |
| 1pgl | Bean Pod Mottle Virus, middle component | Secoviridae | Comovirus |
| 5a32 | Cowpea Mosaic Virus, top component | Secoviridae | Comovirus |
| 2bfu | Cowpea Mosaic Virus, top component | Secoviridae | Comovirus |
| 2xpj | Physalis Mottle Virus | Tymoviridae | Tymovirus |
| 1e57 | Physalis Mottle Virus, empty | Tymoviridae | Tymovirus |
| 2wws | Physalis Mottle Virus, empty | Tymoviridae | Tymovirus |
| 1qjz | Physalis Mottle Virus | Tymoviridae | Tymovirus |
| 2fz2 | Turnip Yellow Mosaic Virus, 100 K | Tymoviridae | Tymovirus |
| 2fz1 | Turnip Yellow Mosaic Virus, 100 K – empty | Tymoviridae | Tymovirus |
| 1auy | Turnip Yellow Mosaic Virus | Tymoviridae | Tymovirus |

Table S1. List of viruses in the dataset (cont'd).

| PDB ID | Virus (entry) | Family | Genus |
|--------|---------------|--------|-------|
| 1qbe | Bacteriophage Q Beta | Leviviridae | Allolevivirus |
| 1gav | Bacteriophage GA | Leviviridae | Levivirus |
| 2w4z | Bacteriophage PhiCb5 | Leviviridae | Levivirus |
| 1frs | Bacteriophage FR | Leviviridae | Levivirus |
| 1dwn | Bacterophage PP7 | Leviviridae | Levivirus |
| 2ms2 | Bacteriophage MS2 | Leviviridae | Levivirus |
| 2vf9 | Bacteriophage PRR1 | Leviviridae | Levivirus |
| 4zor | Phage MS2, mutant | Leviviridae | Levivirus |
| 2buk | Satellite Tobacco Necrosis Virus | Satellites | Satellites |
| 1stm | Satellite Panicum Mosaic Virus | Satellites | Satellites |
| 4nia | Satellite Tobacco Mosaic Virus | Satellites | Satellites |
| 4oq8 | Satellite Tobacco Mosaic Virus | Satellites | Satellites |

Table S1. List of viruses in the dataset (cont'd).

| PDB ID | $\mathcal{T}$ | nr. of chains | avg. capsid radius [nm] | genome length [nt] |
|--------|------|------|------|------|
| 1auy | 3 | 3 | 15.2 | 6318 |
| 1aym | p3 | 4 | 15.5 | 7124 |
| 1b35 | p3 | 4 | 16.0 | 9185 |
| 1bev | p3 | 4 | 15.4 | 7414 |
| 1c8n | 3 | 3 | 15.5 | 3682 |
| 1cov | p3 | 4 | 15.7 | 7389 |
| 1cwp | 3 | 3 | 14.1 | 3171, 2774 |
| 1dwn | 3 | 3 | 14.6 | 3588 |
| 1e57 | 3 | 3 | 15.6 | 6673 |
| 1f15 | 3 | 3 | 14.9 | 3363, 3049 |
| 1f2n | 3 | 3 | 15.0 | 4450 |
| 1f8v | 3 | 3* | 17.2 | 3011, 1311 (4322) |
| 1fmd | p3 | 4 | 14.9 | 8110 |
| 1fpn | p3 | 4 | 15.7 | 7102 |
| 1frs | 3 | 3 | 13.8 | 3575 |
| 1gav | 3 | 1 | 13.9 | 3466 |
| 1h8t | p3 | 4 | 15.7 | 7436 |
| 1hxs | p3 | 4 | 16.0 | 7440 |
| 1ihm | 3 | 3 | 19.9 | 7598 |
| 1js9 | 3 | 3 | 14.1 | 3234, 2867 |
| 1laj | 3 | 3 | 14.9 | 3410, 3074 |
| 1mqt | p3 | 4 | 15.7 | 7401 |
| 1nd2 | p3 | 4 | 15.6 | 7124 |
| 1ng0 | 3 | 3 | 15.2 | 4082 |
| 1nov | 3 | 3* | 17.4 | 3204, 1336 (4540) |
| 1oop | p3 | 4 | 15.7 | 7401 |
| 1opo | 3 | 3 | 17.6 | 3966 |
| 1pgl | p3 | 2 | 15.3 | 5995 |
| 1pgw | p3 | 2 | 15.3 | 3657 |
| 1qbe | 3 | 3 | 14.3 | 4215 |
| 1qjz | 3 | 3 | 15.0 | 6673 |
| 1r1a | p3 | 4 | 15.6 | 7095 |
| 1smv | 3 | 3 | 15.0 | 4148 |
| 1stm | 1 | 1 | 8.4 | 826 |
| 1tme | p3 | 4 | 15.5 | 8093 |
| 1tmf | p3 | 4 | 15.9 | 8093 |
| 1x33 | 3 | 3 | 15.5 | 4148 |
| 1x35 | 3 | 3 | 14.9 | 4148 |
| 1z7s | p3 | 4 | 15.9 | 7406 |
| 1za7 | 3 | 3 | 14.0 | 3171, 2774 |
| 2bbv | 3 | 3* | 16.6 | 3106, 1399 (4505) |
| 2bfu | p3 | 2 | 15.1 | 5889 |
| 2buk | 1 | 1 | 9.6 | 1239 |
| 2fz1 | 3 | 3 | 15.2 | 6318 |
| 2fz2 | 3 | 3 | 15.2 | 6318 |
| 2gh8 | 3 | 3 | 20.6 | 8284 |
| 2izw | 3 | 3 | 15.2 | 4212 |

Table S2. Properties of viruses in the dataset. Shown are the viruses' triangulation ($\mathcal{T}$) number and the average radius of its capsid (both obtained from VIPERdb [2]) as well as the characteristic length of the viral genome (obtained from NCBI Nucleotide database [3]). For viruses from Bromoviridae and Nodaviridae family we list the lengths of the RNA1 and RNA2 segments. As viruses from Nodaviridae family pack both segments into the same virion, we list in parentheses also the total length of both segments. In addition, we list the number of different protein chains used in the averaging procedures for obtaining the viral CP and N-tail properties. In viruses from Nodaviridae family (marked with an asterisk) we use only the longer parts of the cleaved capsid proteins deposited in the database.

| PDB ID | $\mathcal{T}$ | nr. of chains | avg. capsid radius [nm] | genome length [nt] |
|--------|------|------|------|------|
| 2mev | p3 | 4 | 15.4 | 7761 |
| 2ms2 | 3 | 3 | 13.8 | 3569 |
| 2tbv | 3 | 3 | 17.3 | 4776 |
| 2vf9 | 3 | 3 | 13.9 | 3573 |
| 2vq0 | 3 | 3 | 14.9 | 4148 |
| 2w4z | 3 | 3 | 13.9 | 3762 |
| 2wff | p3 | 4 | 15.4 | 7681 |
| 2ws9 | p3 | 4 | 15.7 | 7681 |
| 2wws | 3 | 3 | 15.2 | 6673 |
| 2wzr | p3 | 4 | 15.0 | 8144 |
| 2x5i | p3 | 4 | 15.6 | 7427 |
| 2xpj | 3 | 3 | 15.5 | 6673 |
| 2zah | 3 | 3 | 17.2 | 4267 |
| 2ztn | 1 | 1 | 13.8 | 7176 |
| 3cji | p3 | 4 | 15.7 | 7310 |
| 3hag | 1 | 1 | 13.5 | 7176 |
| 3j1p | 3 | 3 | 21.6 | 7437 |
| 3j27 | p3 | 6 | 23.0 | 10723 |
| 3j7l | 3 | 3 | 14.1 | 3234, 2867 |
| 3jb4 | p3 | 3 | 15.1 | 7590 |
| 3jb8 | 3 | 3 | 16.4 | 4437 |
| 3m8l | 3 | 3 | 20.5 | 7690 |
| 3nap | p3 | 3 | 16.4 | 9010 |
| 3vbs | p3 | 4 | 15.5 | 7405 |
| 3vbu | p3 | 3 | 16.1 | 7405 |
| 3zx8 | 3 | 3 | 17.2 | 4054 |
| 3zx9 | 3 | 3 | 18.7 | 4054 |
| 4aed | p3 | 4 | 15.3 | 7405 |
| 4biq | p3 | 3 | 16.1 | 7403 |
| 4fsj | 3 | 3* | 16.8 | 3107, 1400 (4507) |
| 4ftb | 3 | 3* | 16.8 | 3107, 1400 (4507) |
| 4gh4 | p3 | 4 | 14.9 | 8202 |
| 4llf | 3 | 3 | 17.3 | 4701 |
| 4nia | 1 | 1 | 8.7 | 1058 |
| 4nwv | 3 | 3 | 18.3 | 3421, 2574 (5995) |
| 4nww | 3 | 3 | 18.4 | 3421, 2574 (5995) |
| 4oq8 | 1 | 1 | 8.7 | 1058 |
| 4qpg | p3 | 3 | 15.1 | 7478 |
| 4qpi | p3 | 3 | 15.0 | 7478 |
| 4sbv | 3 | 3 | 15.1 | 4132 |
| 4wm8 | p3 | 4 | 15.7 | 7321 |
| 4z92 | p3 | 3 | 15.0 | 7319 |
| 4zor | 1 | 1 | 8.8 | N/A |
| 5a32 | p3 | 2 | 15.2 | 5889 |
| 5a8f | p3 | 3 | 16.0 | 7991 |
| 5aoo | p3 | 3 | 15.2 | 8251 |
| 5c4w | p3 | 4 | 15.5 | 7413 |
| 5cdc | p3 | 3 | 16.6 | 9499 |
| 5cfc | p3 | 4 | 15.3 | 7991 |
| 5g52 | p3 | 3 | 17.8 | 10140 |
| 5gka | p3 | 3 | 15.5 | 8251 |
| 5ire | p3 | 6 | 23.1 | 10806 |

Table S2. Properties of viruses in the dataset (cont'd).

| PDB ID | $\mathcal{T}$ | nr. of chains | avg. capsid radius [nm] | genome length [nt] |
|--------|------|-----|------|-------|
| 5iz7 | p3 | 6 | 23.0 | 10806 |
| 5j96 | p3 | 3 | 18.7 | 9505 |
| 5j98 | p3 | 3 | 18.0 | 9505 |
| 5jzg | p3 | 3 | 15.5 | 7099 |
| 5k0u | p3 | 4 | 15.4 | 7099 |
| 5l7o | p3 | 3 | 16.1 | 9010 |
| 5lk7 | p3 | 3 | 15.9 | 9505 |
| 5lk8 | p3 | 3 | 16.1 | 9505 |
| 5lvc | p3 | 3 | 15.5 | 8251 |
| 5lwg | p3 | 4 | 16.1 | 9499 |
| 5lwi | p3 | 3 | 16.8 | 9499 |
| 5mqc | p3 | 3 | 16.6 | 8358 |
| 5ned | p3 | 4 | 14.9 | 8241 |
| 5u4w | p3 | 12 | 27.6 | 10806 |
| 5wsn | p3 | 6 | 23.7 | 10970 |
| 5wte | p3 | 3 | 15.2 | 7478 |
| 5wtf | p3 | 3 | 15.4 | 7478 |

Table S2. Properties of viruses in the dataset (cont'd).

| PDB ID | $L_{\rm p}$ [AA] | $L_{\rm t}$ [AA] | $q_{\rm c}\|_{pH=4.5}$ [$e_0$] | $q_{\rm c}\|_{pH=7}$ [$e_0$] | $q_{\rm c}\|_{pH=9.5}$ [$e_0$] | $q_{\rm t}\|_{pH=4.5}$ [$e_0$] | $q_{\rm t}\|_{pH=4.5}$ [$e_0$] | $q_{\rm t}\|_{pH=4.5}$ [$e_0$] |
|---|---|---|---|---|---|---|---|---|
| 1auy | 190.0 | 14.0 | 3.13 | -0.76 | -1.78 | 0.26 | -0.33 | -0.78 |
| 1aym | 213.0 | 13.2 | 4.27 | -2.32 | -3.92 | 0.75 | 0.27 | -0.30 |
| 1b35 | 213.5 | 9.2 | 2.65 | -3.67 | -5.32 | 1.54 | 1.02 | 0.56 |
| 1bev | 209.8 | 30.5 | 0.84 | -3.09 | -4.02 | 0.99 | -0.18 | -1.00 |
| 1c8n | 276.0 | 77.3 | 0.44 | -1.33 | -2.74 | 13.04 | 11.67 | 10.88 |
| 1cov | 212.5 | 20.8 | 2.54 | -1.55 | -3.13 | -0.40 | -1.00 | -1.49 |
| 1cwp | 190.0 | 49.0 | -0.36 | -4.23 | -5.39 | 11.31 | 11.00 | 10.28 |
| 1dwn | 127.0 | 2.0 | 3.80 | 0.11 | -1.00 | 2.00 | 2.00 | 1.55 |
| 1e57 | 188.0 | 23.3 | 3.44 | 0.77 | -0.40 | 0.32 | -0.33 | -0.80 |
| 1f15 | 218.0 | 42.7 | 7.75 | 4.13 | 1.94 | 9.68 | 8.70 | 8.17 |
| 1f2n | 238.0 | 49.3 | 7.03 | 3.53 | 1.90 | 12.98 | 12.66 | 11.94 |
| 1f8v | 197.5 | 44.3 | 2.10 | -0.33 | -0.95 | 14.33 | 14.33 | 13.67 |
| 1fmd | 182.5 | 24.5 | 5.81 | -1.00 | -2.68 | 0.77 | 0.25 | -0.45 |
| 1fpn | 213.8 | 21.5 | 5.78 | -0.35 | -1.98 | 0.02 | -0.73 | -1.30 |
| 1frs | 129.0 | 6.0 | 3.26 | 1.67 | 0.41 | -0.38 | -1.00 | -1.39 |
| 1gav | 129.0 | 5.0 | 2.63 | 1.00 | -0.45 | 2.00 | 2.00 | 1.61 |
| 1h8t | 215.0 | 20.8 | 2.99 | -2.17 | -3.78 | -0.65 | -1.25 | -1.69 |
| 1hxs | 219.8 | 14.5 | 4.76 | -0.92 | -2.53 | 0.04 | -0.25 | -0.74 |
| 1ihm | 530.0 | 36.0 | 7.60 | -4.49 | -6.36 | -1.33 | -2.00 | -2.45 |
| 1js9 | 189.0 | 48.7 | 4.51 | -0.16 | -1.43 | 10.31 | 10.00 | 9.34 |
| 1laj | 217.0 | 38.7 | 10.12 | 4.30 | 2.62 | 8.47 | 8.33 | 7.86 |
| 1mqt | 212.5 | 21.0 | 2.25 | -1.80 | -3.36 | -0.40 | -1.00 | -1.44 |
| 1nd2 | 213.0 | 13.2 | 4.28 | -2.10 | -3.74 | 0.75 | 0.27 | -0.30 |
| 1ng0 | 253.0 | 58.0 | 2.08 | -3.03 | -4.24 | 13.74 | 13.33 | 12.63 |
| 1nov | 199.5 | 37.7 | 3.22 | -1.56 | -2.87 | 14.33 | 14.33 | 13.81 |
| 1oop | 212.8 | 29.0 | 0.98 | -2.60 | -3.88 | 0.19 | -0.95 | -1.62 |
| 1opo | 348.0 | 82.3 | 5.34 | -0.55 | -2.42 | 13.07 | 12.00 | 10.95 |
| 1pgl | 277.5 | 6.0 | 5.68 | -0.74 | -2.05 | 0.65 | 0.50 | 0.11 |
| 1pgw | 268.0 | 4.5 | 5.45 | -0.74 | -1.99 | 1.00 | 1.00 | 0.61 |
| 1qbe | 132.0 | 5.0 | 4.77 | 3.00 | 1.80 | 1.31 | 1.00 | 0.55 |
| 1qjz | 188.0 | 22.0 | 3.76 | 2.00 | 0.96 | 0.32 | -0.33 | -0.80 |
| 1r1a | 208.0 | 24.8 | 3.77 | -2.35 | -3.66 | 1.14 | 0.30 | -0.42 |
| 1smv | 266.0 | 69.3 | 2.28 | -0.23 | -1.14 | 13.98 | 13.67 | 13.06 |
| 1stm | 157.0 | 20.0 | 2.08 | -0.89 | -1.55 | 8.00 | 8.00 | 7.53 |
| 1tme | 212.0 | 17.5 | 2.14 | -1.12 | -2.28 | 0.77 | 0.02 | -0.59 |
| 1tmf | 205.0 | 20.8 | 2.16 | -1.10 | -2.47 | 0.18 | -0.97 | -1.55 |
| 1x33 | 268.0 | 71.0 | 2.39 | 0.10 | -0.87 | 14.98 | 14.67 | 14.00 |
| 1x35 | 268.0 | 71.7 | 0.16 | -2.23 | -3.19 | 14.98 | 14.67 | 14.00 |
| 1z7s | 219.5 | 22.8 | 3.70 | -1.02 | -2.41 | 0.01 | -0.73 | -1.20 |
| 1za7 | 189.0 | 48.0 | -3.00 | -6.56 | -7.79 | 11.31 | 11.00 | 10.34 |
| 2bbv | 203.5 | 60.0 | 0.27 | -3.23 | -3.90 | 19.00 | 19.00 | 18.51 |
| 2bfu | 279.0 | 21.0 | 3.79 | -3.64 | -5.37 | -0.41 | -1.00 | -1.52 |
| 2buk | 196.0 | 13.0 | 7.27 | 2.30 | 1.40 | 5.00 | 5.00 | 4.48 |
| 2fz1 | 189.0 | 13.0 | 2.23 | -1.46 | -2.52 | 0.26 | -0.33 | -0.78 |
| 2fz2 | 189.0 | 13.3 | 2.52 | -1.13 | -2.12 | 0.26 | -0.33 | -0.78 |
| 2gh8 | 558.0 | 14.0 | 13.43 | -5.65 | -10.70 | -1.01 | -1.67 | -2.05 |
| 2izw | 234.0 | 57.3 | 5.82 | 2.44 | 1.65 | 13.74 | 13.33 | 12.09 |
| 2mev | 208.5 | 14.2 | 2.91 | -0.62 | -2.14 | 0.14 | -0.72 | -1.20 |
| 2ms2 | 129.0 | 6.0 | 1.90 | 0.00 | -1.19 | 1.00 | 1.00 | 0.61 |
| 2tbv | 387.0 | 99.7 | 1.89 | -2.22 | -3.17 | 15.61 | 14.43 | 13.24 |

Table S3. Results obtained using the definition of N-tails based on STRIDE secondary structure assignment. Shown are the length of the full capsid protein ($L_{\rm p}$) and the predicted tail length ($L_{\rm t}$), as well as the average charge on the CP ($q_{\rm c}$) and the tail ($q_{\rm t}$) at three different values of $pH$ = 4.5, 7, 9.5. The values are averaged over the chains in a database entry (cf. Table S2).

| PDB ID | $L_p$ [AA] | $L_t$ [AA] | $q_c|_{pH=4.5}$ [$e_0$] | $q_c|_{pH=7}$ [$e_0$] | $q_c|_{pH=9.5}$ [$e_0$] | $q_t|_{pH=4.5}$ [$e_0$] | $q_t|_{pH=4.5}$ [$e_0$] | $q_t|_{pH=4.5}$ [$e_0$] |
|---|---|---|---|---|---|---|---|---|
| 2vf9 | 131.0 | 5.0 | 4.46 | -0.13 | -1.44 | 1.00 | 1.00 | 0.61 |
| 2vq0 | 256.0 | 58.3 | 1.78 | -0.56 | -1.43 | 14.98 | 14.67 | 14.00 |
| 2w4z | 122.0 | 4.0 | 6.35 | 0.64 | -1.46 | 0.14 | -0.00 | -0.39 |
| 2wff | 195.5 | 35.5 | 5.37 | -0.10 | -1.20 | -0.68 | -2.20 | -2.79 |
| 2ws9 | 195.5 | 29.2 | 5.07 | -1.05 | -2.30 | 0.17 | -1.42 | -2.04 |
| 2wws | 188.0 | 13.7 | 6.47 | 4.37 | 3.03 | -0.12 | -0.67 | -1.10 |
| 2wzr | 186.0 | 22.0 | 3.92 | -2.82 | -4.71 | 1.15 | 0.52 | -0.24 |
| 2x5i | 215.0 | 20.8 | 4.37 | -1.40 | -3.33 | -0.50 | -1.23 | -1.69 |
| 2xpj | 188.0 | 13.3 | 6.08 | 4.04 | 2.85 | -0.12 | -0.67 | -1.10 |
| 2zah | 390.0 | 92.3 | 4.76 | -0.70 | -2.80 | 15.32 | 14.66 | 13.23 |
| 2ztn | 478.0 | 3.0 | 10.48 | 0.41 | -2.65 | 1.00 | 1.00 | 0.61 |
| 3cji | 214.2 | 18.2 | 3.72 | -0.82 | -2.44 | 0.05 | -0.73 | -1.31 |
| 3hag | 504.0 | 21.0 | 17.39 | 0.20 | -3.91 | -0.58 | -1.00 | -1.39 |
| 3j1p | 579.0 | 47.7 | -3.42 | -9.22 | -11.72 | -0.79 | -2.00 | -2.45 |
| 3j27 | 285.0 | 16.5 | 11.08 | 0.10 | -2.13 | 1.70 | 0.73 | 0.27 |
| 3j7l | 189.0 | 48.0 | 3.41 | -1.79 | -3.14 | 10.31 | 10.00 | 9.41 |
| 3jb4 | 266.7 | 37.3 | 2.69 | -4.89 | -6.54 | 0.49 | -1.30 | -1.89 |
| 3jb8 | 236.0 | 46.0 | 5.77 | 3.10 | 2.26 | 12.31 | 12.00 | 11.46 |
| 3m8l | 534.0 | 27.0 | 9.80 | -6.18 | -10.53 | -1.08 | -2.00 | -2.41 |
| 3nap | 270.3 | 15.3 | 0.13 | -5.82 | -7.51 | 1.43 | 1.33 | 0.84 |
| 3vbs | 210.5 | 17.5 | 1.83 | -1.85 | -2.92 | 1.13 | 0.07 | -0.57 |
| 3vbu | 233.7 | 9.7 | 2.65 | -3.33 | -4.92 | 0.62 | 0.03 | -0.41 |
| 3zx8 | 347.0 | 105.0 | 5.44 | -0.22 | -1.82 | 13.79 | 12.00 | 10.56 |
| 3zx9 | 347.0 | 105.0 | 5.18 | -1.22 | -2.89 | 13.79 | 12.00 | 10.56 |
| 4aed | 215.5 | 23.8 | 1.45 | -2.57 | -3.71 | 1.29 | 0.07 | -0.62 |
| 4biq | 220.0 | 6.7 | 3.84 | -2.00 | -4.54 | 0.69 | 0.33 | -0.12 |
| 4fsj | 203.5 | 60.0 | -0.91 | -4.56 | -5.32 | 19.00 | 19.00 | 18.57 |
| 4ftb | 203.5 | 60.0 | -1.20 | -4.89 | -5.65 | 19.00 | 19.00 | 18.57 |
| 4gh4 | 170.8 | 18.2 | 6.50 | 0.40 | -1.26 | 0.75 | 0.50 | -0.21 |
| 4llf | 380.0 | 89.7 | 0.85 | -4.14 | -4.64 | 15.75 | 14.43 | 12.86 |
| 4nia | 159.0 | 20.0 | 3.66 | 2.01 | 1.73 | 5.14 | 5.00 | 4.42 |
| 4nwv | 391.0 | 46.7 | 7.09 | -2.73 | -5.88 | 13.94 | 12.19 | 11.16 |
| 4nww | 358.0 | 6.7 | 12.62 | -2.77 | -6.54 | 1.00 | 1.00 | 0.61 |
| 4oq8 | 159.0 | 36.0 | 5.27 | 3.01 | 2.58 | 7.14 | 7.00 | 6.15 |
| 4qpg | 225.0 | 19.3 | 7.64 | 0.20 | -2.02 | 1.49 | 0.40 | -0.16 |
| 4qpi | 248.7 | 24.3 | 6.50 | -0.83 | -3.12 | 0.56 | -0.33 | -0.79 |
| 4sbv | 260.0 | 63.3 | 4.79 | 2.10 | 1.40 | 15.09 | 15.00 | 14.21 |
| 4wm8 | 215.0 | 21.8 | 2.55 | -3.15 | -4.26 | 1.17 | -0.15 | -0.85 |
| 4z92 | 258.7 | 34.3 | 2.29 | -4.30 | -6.20 | 1.38 | 0.33 | -0.35 |
| 4zor | 129.0 | 6.0 | 2.24 | 0.20 | -0.99 | 1.00 | 1.00 | 0.61 |
| 5a32 | 279.0 | 7.5 | 2.76 | -4.64 | -6.30 | -0.12 | -0.50 | -0.99 |
| 5a8f | 222.0 | 3.3 | 8.23 | 3.93 | 2.32 | 1.09 | 1.00 | 0.54 |
| 5aoo | 282.0 | 19.7 | 0.42 | -3.86 | -4.99 | 0.38 | -0.30 | -0.81 |
| 5c4w | 215.5 | 27.2 | 1.63 | -2.30 | -3.46 | 0.78 | -0.43 | -1.12 |
| 5cdc | 248.7 | 27.3 | -1.51 | -6.22 | -8.04 | 0.54 | -0.93 | -1.59 |
| 5cfc | 191.5 | 11.5 | 4.79 | 1.43 | 0.07 | 0.97 | 0.52 | -0.04 |
| 5g52 | 296.7 | 17.0 | 5.10 | -1.33 | -3.60 | 0.59 | -0.60 | -1.12 |
| 5gka | 280.7 | 59.7 | 0.19 | -3.80 | -5.10 | 1.50 | 0.03 | -0.70 |
| 5ire | 289.5 | 7.0 | 10.80 | -0.14 | -2.39 | 1.50 | 1.50 | 1.11 |
| 5iz7 | 289.5 | 13.0 | 10.83 | 0.76 | -1.31 | 2.14 | 1.70 | 1.27 |

Table S3. Results obtained using the definition of N-tails based on STRIDE secondary structure assignment (cont'd).

| PDB ID | $L_p$ [Å] | $L_t$ [Å] | $q_c\|_{pH=4.5}$ [$e_0$] | $q_c\|_{pH=7}$ [$e_0$] | $q_c\|_{pH=9.5}$ [$e_0$] | $q_t\|_{pH=4.5}$ [$e_0$] | $q_t\|_{pH=4.5}$ [$e_0$] | $q_t\|_{pH=4.5}$ [$e_0$] |
|---|---|---|---|---|---|---|---|---|
| 5j96 | 319.0 | 17.7 | 1.10 | -6.09 | -8.38 | 0.97 | -0.27 | -0.74 |
| 5j98 | 319.0 | 13.3 | 1.87 | -5.12 | -7.41 | 0.53 | -0.60 | -1.08 |
| 5jzg | 281.7 | 64.0 | 8.51 | 1.50 | -0.32 | 1.56 | -1.20 | -2.32 |
| 5k0u | 211.2 | 39.2 | 5.42 | 0.10 | -1.36 | 1.83 | 0.10 | -0.72 |
| 5l7o | 270.3 | 47.7 | 0.86 | -3.86 | -5.66 | 1.26 | -0.90 | -1.61 |
| 5lk7 | 319.0 | 15.7 | 2.18 | -4.82 | -7.41 | 0.64 | -0.60 | -1.08 |
| 5lk8 | 319.0 | 28.7 | 2.86 | -4.76 | -7.50 | 0.32 | -1.27 | -1.83 |
| 5lvc | 282.0 | 30.0 | -1.63 | -5.53 | -6.50 | -0.04 | -0.97 | -1.56 |
| 5lwg | 203.0 | 36.8 | -0.66 | -4.17 | -5.40 | 1.25 | -0.42 | -1.10 |
| 5lwi | 248.0 | 14.0 | -0.66 | -6.49 | -8.27 | 1.59 | 1.33 | 0.75 |
| 5mqc | 259.3 | 15.0 | 1.22 | -5.09 | -7.15 | 0.74 | 0.33 | -0.16 |
| 5ned | 183.5 | 24.8 | 5.92 | -0.80 | -2.48 | 0.77 | 0.25 | -0.40 |
| 5u4w | 150.5 | 2.5 | 4.98 | -1.41 | -2.52 | 0.90 | 0.75 | 0.26 |
| 5wsn | 287.0 | 14.0 | 12.20 | 2.76 | 0.66 | 1.29 | 0.55 | 0.08 |
| 5wte | 248.7 | 26.7 | 6.78 | -0.80 | -3.06 | 0.71 | -0.33 | -0.88 |
| 5wtf | 224.3 | 18.7 | 7.62 | -0.10 | -2.37 | 1.49 | 0.40 | -0.16 |

Table S3. Results obtained using the definition of N-tails based on STRIDE secondary structure assignment (cont'd).

| PDB ID | $L_p$ [AA] | $L_t$ [AA] | $q_c|_{pH=4.5}$ [$e_0$] | $q_c|_{pH=7}$ [$e_0$] | $q_c|_{pH=9.5}$ [$e_0$] | $q_t|_{pH=4.5}$ [$e_0$] | $q_t|_{pH=4.5}$ [$e_0$] | $q_t|_{pH=4.5}$ [$e_0$] |
|---|---|---|---|---|---|---|---|---|
| 1auy | 190.0 | 17.0 | 3.50 | -0.10 | -1.12 | -0.10 | -1.00 | -1.45 |
| 1aym | 213.0 | 8.8 | 3.84 | -2.82 | -4.49 | 1.18 | 0.77 | 0.26 |
| 1b35 | 213.5 | 13.5 | 3.09 | -2.92 | -4.55 | 1.10 | 0.28 | -0.20 |
| 1bev | 209.8 | 11.0 | 0.41 | -4.29 | -5.59 | 1.42 | 1.02 | 0.57 |
| 1c8n | 276.0 | 15.0 | 11.03 | 8.34 | 6.65 | 2.45 | 2.00 | 1.48 |
| 1cov | 212.5 | 10.0 | 1.59 | -2.57 | -4.21 | 0.76 | 0.27 | -0.15 |
| 1cwp | 190.0 | 26.0 | 0.95 | -3.22 | -4.51 | 10.00 | 10.00 | 9.41 |
| 1dwn | 127.0 | 7.0 | 3.80 | 0.11 | -1.00 | 2.00 | 2.00 | 1.55 |
| 1e57 | 188.0 | 27.0 | 3.01 | 0.44 | -0.69 | 0.76 | 0.00 | -0.51 |
| 1f15 | 218.0 | 33.0 | 8.84 | 4.83 | 2.58 | 8.59 | 8.00 | 7.53 |
| 1f2n | 238.0 | 50.0 | 6.70 | 3.20 | 1.57 | 13.31 | 13.00 | 12.28 |
| 1f8v | 197.5 | 44.0 | 2.77 | 0.34 | -0.28 | 14.00 | 14.00 | 13.34 |
| 1fmd | 182.5 | 12.8 | 5.68 | -1.03 | -2.96 | 0.91 | 0.27 | -0.17 |
| 1fpn | 213.8 | 8.0 | 4.87 | -1.60 | -3.35 | 1.15 | 0.77 | 0.31 |
| 1frs | 129.0 | 3.0 | 1.87 | -0.33 | -1.59 | 1.00 | 1.00 | 0.61 |
| 1gav | 129.0 | 0.0 | 3.63 | 2.00 | 0.54 | 1.00 | 1.00 | 0.61 |
| 1h8t | 215.0 | 9.5 | 2.13 | -2.95 | -4.51 | 0.41 | -0.23 | -0.70 |
| 1hxs | 219.8 | 22.0 | 3.79 | -1.22 | -2.84 | 1.05 | 0.05 | -0.42 |
| 1ihm | 530.0 | 34.0 | 8.41 | -3.49 | -5.36 | -2.13 | -3.00 | -3.45 |
| 1js9 | 189.0 | 23.0 | 6.82 | 1.84 | 0.37 | 8.00 | 8.00 | 7.53 |
| 1laj | 217.0 | 32.0 | 10.45 | 4.64 | 2.95 | 8.14 | 8.00 | 7.53 |
| 1mqt | 212.5 | 19.0 | 1.65 | -2.32 | -3.89 | 0.19 | -0.48 | -0.90 |
| 1nd2 | 213.0 | 8.8 | 3.85 | -2.60 | -4.30 | 1.18 | 0.77 | 0.26 |
| 1ng0 | 253.0 | 57.0 | 0.50 | -4.70 | -5.88 | 15.31 | 15.00 | 14.27 |
| 1nov | 199.5 | 49.0 | -0.45 | -5.22 | -6.57 | 18.00 | 18.00 | 17.51 |
| 1oop | 212.8 | 16.2 | 0.77 | -3.32 | -4.85 | 0.41 | -0.23 | -0.65 |
| 1opo | 348.0 | 85.0 | 5.34 | -0.55 | -2.42 | 13.07 | 12.00 | 10.95 |
| 1pgl | 277.5 | 9.5 | 6.38 | 0.26 | -1.02 | -0.05 | -0.50 | -0.92 |
| 1pgw | 268.0 | 5.5 | 5.38 | -0.74 | -1.95 | 1.07 | 1.00 | 0.58 |
| 1qbe | 132.0 | 7.0 | 4.77 | 3.00 | 1.80 | 1.31 | 1.00 | 0.55 |
| 1qjz | 188.0 | 27.0 | 3.32 | 1.67 | 0.67 | 0.76 | 0.00 | -0.51 |
| 1r1a | 208.0 | 9.2 | 3.76 | -2.82 | -4.39 | 1.15 | 0.77 | 0.31 |
| 1smv | 266.0 | 60.0 | 3.94 | 1.44 | 0.52 | 12.31 | 12.00 | 11.40 |
| 1stm | 157.0 | 18.0 | 2.08 | -0.89 | -1.55 | 8.00 | 8.00 | 7.53 |
| 1tme | 212.0 | 11.2 | 2.29 | -1.35 | -2.68 | 0.02 | -0.50 | -0.93 |
| 1tmf | 205.0 | 7.5 | 2.31 | -1.57 | -3.05 | 0.24 | -0.25 | -0.67 |
| 1x33 | 268.0 | 63.0 | 3.06 | 0.77 | -0.21 | 14.31 | 14.00 | 13.33 |
| 1x35 | 268.0 | 70.0 | 0.83 | -1.56 | -2.52 | 14.31 | 14.00 | 13.33 |
| 1z7s | 219.5 | 17.5 | 2.84 | -2.02 | -3.45 | 0.87 | 0.27 | -0.17 |
| 1za7 | 189.0 | 25.0 | -1.69 | -5.56 | -6.86 | 10.00 | 10.00 | 9.41 |
| 2bbv | 203.5 | 57.0 | 1.27 | -2.23 | -2.90 | 18.00 | 18.00 | 17.51 |
| 2bfu | 279.0 | 6.0 | 3.07 | -4.64 | -6.51 | 0.31 | -0.00 | -0.39 |
| 2buk | 196.0 | 16.0 | 6.27 | 1.30 | 0.40 | 6.00 | 6.00 | 5.48 |
| 2fz1 | 189.0 | 16.0 | 2.60 | -0.80 | -1.85 | -0.10 | -1.00 | -1.45 |
| 2fz2 | 189.0 | 16.0 | 2.89 | -0.46 | -1.45 | -0.10 | -1.00 | -1.45 |
| 2gh8 | 558.0 | 22.0 | 14.36 | -4.32 | -9.37 | -1.93 | -3.00 | -3.39 |
| 2izw | 234.0 | 57.0 | 5.24 | 1.77 | 0.85 | 14.31 | 14.00 | 12.89 |
| 2mev | 208.5 | 14.8 | 3.63 | -0.10 | -1.68 | -0.75 | -1.50 | -1.92 |
| 2ms2 | 129.0 | 1.0 | 1.90 | 0.00 | -1.19 | 1.00 | 1.00 | 0.61 |
| 2tbv | 387.0 | 15.0 | 14.50 | 9.21 | 7.53 | 3.00 | 3.00 | 2.55 |

Table S4. Results obtained using the definition of N-tails based on MD2 prediction of intrinsically disordered regions in proteins. Shown are the length of the full capsid protein ($L_p$) and the predicted tail length ($L_t$), as well as the average charge on the CP ($q_c$) and the tail ($q_t$) at three different values of $pH = 4.5, 7, 9.5$. The values are averaged over the chains in a database entry (cf. Table S2).

| PDB ID | $L_p$ [AA] | $L_t$ [AA] | $q_c|_{pH=4.5}$ [$e_0$] | $q_c|_{pH=7}$ [$e_0$] | $q_c|_{pH=9.5}$ [$e_0$] | $q_t|_{pH=4.5}$ [$e_0$] | $q_t|_{pH=4.5}$ [$e_0$] | $q_t|_{pH=4.5}$ [$e_0$] |
|---|---|---|---|---|---|---|---|---|
| 2vf9 | 131.0 | 7.0 | 4.46 | -0.13 | -1.44 | 1.00 | 1.00 | 0.61 |
| 2vq0 | 256.0 | 61.0 | 1.45 | -0.90 | -1.76 | 15.31 | 15.00 | 14.33 |
| 2w4z | 122.0 | 2.0 | 5.49 | -0.36 | -2.46 | 1.00 | 1.00 | 0.61 |
| 2wff | 195.5 | 28.5 | 4.48 | -1.57 | -2.70 | 0.21 | -0.72 | -1.29 |
| 2ws9 | 195.5 | 29.0 | 5.03 | -1.75 | -3.05 | 0.21 | -0.72 | -1.29 |
| 2wws | 188.0 | 27.0 | 5.60 | 3.70 | 2.45 | 0.76 | 0.00 | -0.51 |
| 2wzr | 186.0 | 15.5 | 4.85 | -1.82 | -4.03 | 0.23 | -0.48 | -0.92 |
| 2x5i | 215.0 | 19.8 | 4.07 | -1.65 | -3.62 | 0.09 | -0.95 | -1.40 |
| 2xpj | 188.0 | 27.0 | 5.21 | 3.37 | 2.27 | 0.76 | 0.00 | -0.51 |
| 2zah | 390.0 | 10.0 | 17.54 | 11.63 | 8.55 | 3.00 | 3.00 | 2.55 |
| 2ztn | 478.0 | 7.0 | 8.48 | -1.59 | -4.64 | 3.00 | 3.00 | 2.61 |
| 3cji | 214.2 | 11.5 | 3.97 | -0.57 | -2.34 | -0.20 | -0.97 | -1.40 |
| 3hag | 504.0 | 15.0 | 17.53 | 0.20 | -3.91 | -0.72 | -1.00 | -1.39 |
| 3j1p | 579.0 | 38.0 | -4.11 | -10.22 | -12.72 | -0.10 | -1.00 | -1.45 |
| 3j27 | 285.0 | 4.0 | 10.45 | -1.05 | -3.31 | 1.50 | 1.50 | 1.11 |
| 3j7l | 189.0 | 23.0 | 5.72 | 0.21 | -1.27 | 8.00 | 8.00 | 7.53 |
| 3jb4 | 266.7 | 36.3 | 3.07 | -4.56 | -6.23 | 0.45 | -1.30 | -1.87 |
| 3jb8 | 236.0 | 43.0 | 5.77 | 3.10 | 2.26 | 12.31 | 12.00 | 11.46 |
| 3m8l | 534.0 | 10.0 | 8.41 | -8.18 | -12.55 | 0.31 | -0.00 | -0.39 |
| 3nap | 270.3 | 16.0 | 0.20 | -5.49 | -7.22 | 1.36 | 1.00 | 0.55 |
| 3vbs | 210.5 | 8.5 | 3.24 | -0.80 | -2.11 | -0.06 | -0.73 | -1.14 |
| 3vbu | 233.7 | 3.3 | 2.41 | -3.66 | -5.28 | 0.86 | 0.37 | -0.05 |
| 3zx8 | 347.0 | 83.0 | 7.43 | 1.12 | -0.88 | 12.04 | 11.00 | 9.95 |
| 3zx9 | 347.0 | 83.0 | 6.94 | -0.21 | -2.28 | 12.04 | 11.00 | 9.95 |
| 4aed | 215.5 | 13.2 | 2.47 | -2.02 | -3.42 | 0.48 | -0.23 | -0.65 |
| 4biq | 220.0 | 3.0 | 4.22 | -1.67 | -4.28 | 0.31 | -0.00 | -0.39 |
| 4fsj | 203.5 | 56.0 | 0.09 | -3.56 | -4.32 | 18.00 | 18.00 | 17.57 |
| 4ftb | 203.5 | 56.0 | -0.20 | -3.89 | -4.66 | 18.00 | 18.00 | 17.57 |
| 4gh4 | 170.8 | 9.0 | 6.46 | 0.63 | -1.28 | 0.79 | 0.27 | -0.19 |
| 4llf | 380.0 | 13.0 | 13.60 | 7.29 | 5.61 | 3.00 | 3.00 | 2.61 |
| 4nia | 159.0 | 19.0 | 3.66 | 2.01 | 1.73 | 5.14 | 5.00 | 4.42 |
| 4nwv | 391.0 | 38.0 | 8.09 | -1.73 | -4.88 | 12.94 | 11.19 | 10.16 |
| 4nww | 358.0 | 5.0 | 12.62 | -2.77 | -6.54 | 1.00 | 1.00 | 0.61 |
| 4oq8 | 159.0 | 19.0 | 7.27 | 5.01 | 4.32 | 5.14 | 5.00 | 4.42 |
| 4qpg | 225.0 | 5.3 | 7.69 | -0.73 | -3.06 | 1.44 | 1.33 | 0.88 |
| 4qpi | 248.7 | 14.0 | 7.69 | 0.17 | -2.19 | -0.63 | -1.33 | -1.72 |
| 4sbv | 260.0 | 51.0 | 6.88 | 4.10 | 3.33 | 13.00 | 13.00 | 12.28 |
| 4wm8 | 215.0 | 8.8 | 2.71 | -3.82 | -5.21 | 1.01 | 0.52 | 0.10 |
| 4z92 | 258.7 | 16.7 | 0.76 | -6.63 | -8.60 | 2.91 | 2.66 | 2.04 |
| 4zor | 129.0 | 1.0 | 2.24 | 0.20 | -0.99 | 1.00 | 1.00 | 0.61 |
| 5a32 | 279.0 | 6.0 | 2.33 | -5.14 | -6.90 | 0.31 | -0.00 | -0.39 |
| 5a8f | 222.0 | 3.3 | 8.56 | 4.27 | 2.65 | 0.76 | 0.67 | 0.28 |
| 5aoo | 282.0 | 61.3 | -0.98 | -4.23 | -5.12 | 1.78 | 0.07 | -0.67 |
| 5c4w | 215.5 | 11.8 | 2.30 | -2.25 | -3.69 | 0.33 | -0.23 | -0.64 |
| 5cdc | 248.7 | 19.7 | -1.72 | -6.86 | -8.86 | 0.75 | -0.30 | -0.76 |
| 5cfc | 191.5 | 11.7 | 8.15 | 3.60 | 1.81 | -0.18 | -0.67 | -1.10 |
| 5g52 | 296.7 | 16.0 | 6.51 | -0.27 | -2.58 | -0.82 | -1.67 | -2.14 |
| 5gka | 280.7 | 59.7 | 0.24 | -3.80 | -5.12 | 1.45 | 0.03 | -0.67 |
| 5ire | 289.5 | 10.0 | 9.80 | -1.14 | -3.35 | 2.99 | 2.55 | 2.08 |
| 5iz7 | 289.5 | 10.0 | 10.07 | -0.07 | -2.12 | 2.99 | 2.55 | 2.08 |

Table S4. Results obtained using the definition of N-tails based on MD2 prediction of intrinsically disordered regions in proteins (cont'd).

| **PDB ID** | $L_\mathrm{p}$ [AA] | $L_\mathrm{t}$ [AA] | $q_\mathrm{c}\vert_{pH=4.5}$ [$e_0$] | $q_\mathrm{c}\vert_{pH=7}$ [$e_0$] | $q_\mathrm{c}\vert_{pH=9.5}$ [$e_0$] | $q_\mathrm{t}\vert_{pH=4.5}$ [$e_0$] | $q_\mathrm{t}\vert_{pH=4.5}$ [$e_0$] | $q_\mathrm{t}\vert_{pH=4.5}$ [$e_0$] |
|---|---|---|---|---|---|---|---|---|
| 5j96 | 319.0 | 9.7 | 1.52 | -6.36 | -8.72 | 0.55 | 0.00 | -0.41 |
| 5j98 | 319.0 | 9.7 | 1.62 | -6.05 | -8.41 | 0.55 | 0.00 | -0.41 |
| 5jzg | 281.7 | 16.7 | 9.35 | 0.27 | -2.25 | 0.73 | 0.03 | -0.39 |
| 5k0u | 211.2 | 12.8 | 6.62 | 0.18 | -1.68 | 0.62 | 0.02 | -0.39 |
| 5l7o | 270.3 | 16.0 | 0.76 | -5.76 | -7.82 | 1.36 | 1.00 | 0.55 |
| 5lk7 | 319.0 | 9.7 | 2.26 | -5.42 | -8.07 | 0.55 | 0.00 | -0.41 |
| 5lk8 | 319.0 | 9.7 | 2.64 | -6.02 | -8.92 | 0.55 | 0.00 | -0.41 |
| 5lvc | 282.0 | 61.3 | -1.87 | -5.23 | -6.17 | 1.78 | 0.07 | -0.67 |
| 5lwg | 203.0 | 17.8 | -0.72 | -4.89 | -6.30 | 1.05 | 0.05 | -0.44 |
| 5lwi | 248.0 | 17.0 | 0.05 | -5.19 | -7.09 | 0.88 | 0.03 | -0.43 |
| 5mqc | 259.3 | 12.7 | 2.06 | -4.09 | -6.17 | 0.23 | -0.33 | -0.81 |
| 5ned | 183.5 | 14.0 | 6.66 | -0.05 | -1.91 | 0.03 | -0.50 | -0.97 |
| 5u4w | 150.5 | 9.2 | 2.92 | -3.70 | -4.86 | 2.38 | 1.57 | 1.05 |
| 5wsn | 287.0 | 7.0 | 12.35 | 2.76 | 0.62 | 1.14 | 0.55 | 0.11 |
| 5wte | 248.7 | 14.0 | 8.12 | 0.20 | -2.21 | -0.63 | -1.33 | -1.72 |
| 5wtf | 224.3 | 5.3 | 7.67 | -1.03 | -3.41 | 1.44 | 1.33 | 0.88 |

Table S4. Results obtained using the definition of N-tails based on MD2 prediction of intrinsically disordered regions in proteins (cont'd).

| **PDB ID** | STRIDE: $N_t^+$ | $N_t^-$ | $N_c^+$ | $N_c^-$ | MD2: $N_t^+$ | $N_t^-$ | $N_c^+$ | $N_c^-$ |
|---|---|---|---|---|---|---|---|---|
| 1auy | 1.3 | 2.7 | 10.0 | 9.7 | 2.0 | 4.0 | 9.3 | 8.3 |
| 1aym | 0.8 | 2.0 | 16.5 | 17.5 | 0.8 | 1.2 | 16.5 | 18.2 |
| 1b35 | 1.5 | 1.2 | 16.8 | 21.0 | 1.8 | 2.2 | 16.5 | 20.0 |
| 1bev | 2.2 | 4.2 | 11.8 | 15.0 | 1.2 | 1.0 | 12.8 | 18.2 |
| 1c8n | 15.7 | 5.0 | 10.3 | 16.3 | 3.0 | 2.0 | 23.0 | 19.3 |
| 1cov | 0.8 | 3.2 | 12.0 | 15.8 | 0.5 | 1.0 | 12.2 | 17.8 |
| 1cwp | 11.0 | 1.0 | 10.7 | 16.0 | 9.0 | 0.0 | 12.7 | 17.0 |
| 1dwn | 1.0 | 0.0 | 15.0 | 16.0 | 1.0 | 0.0 | 15.0 | 16.0 |
| 1e57 | 1.3 | 2.7 | 10.3 | 11.3 | 2.0 | 3.0 | 9.7 | 11.0 |
| 1f15 | 11.3 | 3.3 | 20.7 | 21.3 | 10.0 | 3.0 | 22.0 | 21.7 |
| 1f2n | 12.7 | 1.0 | 15.7 | 14.7 | 13.0 | 1.0 | 15.3 | 14.7 |
| 1f8v | 13.3 | 0.0 | 13.7 | 14.7 | 13.0 | 0.0 | 14.3 | 14.7 |
| 1fmd | 1.8 | 3.8 | 16.0 | 15.5 | 1.2 | 1.8 | 16.5 | 17.5 |
| 1fpn | 0.8 | 3.0 | 16.5 | 16.5 | 0.5 | 0.8 | 16.8 | 18.5 |
| 1frs | 0.0 | 2.0 | 10.0 | 11.3 | 0.0 | 0.0 | 10.0 | 13.3 |
| 1gav | 1.0 | 0.0 | 10.0 | 13.0 | 0.0 | 0.0 | 11.0 | 13.0 |
| 1h8t | 0.5 | 3.0 | 13.2 | 16.0 | 0.5 | 1.8 | 13.2 | 17.0 |
| 1hxs | 0.2 | 2.0 | 16.8 | 17.8 | 1.8 | 2.2 | 15.2 | 17.5 |
| 1ihm | 1.0 | 4.0 | 29.3 | 28.7 | 1.0 | 5.0 | 29.3 | 27.7 |
| 1js9 | 10.0 | 1.3 | 15.3 | 15.7 | 7.0 | 0.0 | 18.3 | 17.0 |
| 1laj | 8.3 | 1.0 | 23.7 | 18.7 | 8.0 | 1.0 | 24.0 | 18.7 |
| 1mqt | 0.8 | 3.0 | 11.8 | 15.5 | 0.8 | 2.0 | 11.8 | 16.5 |
| 1nd2 | 0.8 | 2.0 | 16.5 | 17.8 | 0.8 | 1.2 | 16.5 | 18.5 |
| 1ng0 | 14.0 | 1.7 | 13.7 | 15.7 | 15.0 | 1.0 | 12.7 | 16.3 |
| 1nov | 13.3 | 0.0 | 18.7 | 21.0 | 17.0 | 0.0 | 15.0 | 21.0 |
| 1oop | 1.8 | 4.2 | 10.2 | 14.5 | 0.8 | 1.8 | 11.2 | 17.0 |
| 1opo | 15.0 | 5.0 | 25.7 | 28.0 | 15.0 | 5.0 | 25.7 | 28.0 |
| 1pgl | 0.0 | 0.5 | 22.0 | 21.0 | 0.5 | 2.0 | 21.5 | 19.5 |
| 1pgw | 0.0 | 0.0 | 21.0 | 20.0 | 0.5 | 0.5 | 20.5 | 19.5 |
| 1qbe | 1.0 | 1.0 | 13.0 | 13.0 | 1.0 | 1.0 | 13.0 | 13.0 |
| 1qjz | 1.3 | 2.7 | 10.3 | 10.7 | 2.0 | 3.0 | 9.7 | 10.3 |
| 1r1a | 1.8 | 3.2 | 15.0 | 15.2 | 0.5 | 0.8 | 16.2 | 17.8 |
| 1smv | 13.7 | 1.0 | 9.3 | 10.7 | 12.0 | 1.0 | 11.0 | 10.7 |
| 1stm | 7.0 | 0.0 | 10.0 | 11.0 | 7.0 | 0.0 | 10.0 | 11.0 |
| 1tme | 1.5 | 3.0 | 12.0 | 14.5 | 1.0 | 2.5 | 12.5 | 15.8 |
| 1tmf | 2.2 | 4.5 | 11.2 | 14.5 | 0.8 | 2.0 | 12.8 | 16.5 |
| 1x33 | 14.7 | 1.0 | 9.3 | 10.7 | 14.0 | 1.0 | 10.0 | 10.7 |
| 1x35 | 14.7 | 1.0 | 7.3 | 11.3 | 14.0 | 1.0 | 8.0 | 11.3 |
| 1z7s | 1.0 | 2.8 | 14.5 | 16.2 | 1.0 | 1.5 | 14.5 | 17.5 |
| 1za7 | 11.0 | 1.0 | 7.3 | 16.0 | 9.0 | 0.0 | 9.3 | 17.0 |
| 2bbv | 18.0 | 0.0 | 12.7 | 14.7 | 17.0 | 0.0 | 13.7 | 14.7 |
| 2bfu | 1.0 | 3.5 | 21.0 | 25.0 | 0.0 | 1.0 | 22.0 | 27.5 |
| 2buk | 4.0 | 0.0 | 18.0 | 13.0 | 5.0 | 0.0 | 17.0 | 13.0 |
| 2fz1 | 1.3 | 2.7 | 9.7 | 10.7 | 2.0 | 4.0 | 9.0 | 9.3 |
| 2fz2 | 1.3 | 2.7 | 9.7 | 10.0 | 2.0 | 4.0 | 9.0 | 8.7 |
| 2gh8 | 0.0 | 2.7 | 50.0 | 57.0 | 0.0 | 4.0 | 50.0 | 55.7 |
| 2izw | 14.0 | 3.3 | 13.0 | 11.3 | 14.0 | 2.0 | 13.0 | 12.7 |
| 2mev | 1.0 | 2.8 | 13.2 | 16.5 | 0.8 | 3.2 | 13.5 | 16.2 |
| 2ms2 | 0.0 | 0.0 | 9.3 | 12.3 | 0.0 | 0.0 | 9.3 | 12.3 |
| 2tbv | 15.3 | 2.0 | 15.7 | 18.7 | 2.0 | 0.0 | 29.0 | 20.7 |

Table S5. Results for the total number $N$ of residues carrying positive ($N^+$) or negative ($N^-$) charge on the N-tails ($N_t$) and CPs ($N_c$) of viruses in the dataset. The charged residues were obtained using the definition of N-tails based either on STRIDE secondary structure assignment or on MD2 prediction of intrinsically disordered regions in proteins. The numbers of charged residues are averaged over the chains in a database entry (cf. Table S2).

| **PDB ID** | STRIDE: $N_t^+$ | $N_t^-$ | $N_c^+$ | $N_c^-$ | MD2: $N_t^+$ | $N_t^-$ | $N_c^+$ | $N_c^-$ |
|---|---|---|---|---|---|---|---|---|
| 2vf9 | 0.0 | 0.0 | 17.0 | 17.7 | 0.0 | 0.0 | 17.0 | 17.7 |
| 2vq0 | 14.7 | 1.0 | 8.7 | 10.3 | 15.0 | 1.0 | 8.3 | 10.3 |
| 2w4z | 0.0 | 1.0 | 16.0 | 18.3 | 0.0 | 0.0 | 16.0 | 19.3 |
| 2wff | 2.0 | 5.2 | 13.5 | 10.5 | 1.5 | 3.5 | 14.0 | 12.2 |
| 2ws9 | 2.2 | 4.5 | 14.2 | 12.2 | 1.5 | 3.5 | 15.0 | 13.2 |
| 2wws | 0.7 | 2.3 | 13.0 | 11.3 | 2.0 | 3.0 | 11.7 | 10.7 |
| 2wzr | 1.8 | 3.5 | 16.5 | 19.8 | 1.0 | 2.2 | 17.2 | 21.0 |
| 2x5i | 0.8 | 3.0 | 15.2 | 18.5 | 1.0 | 2.5 | 15.2 | 19.0 |
| 2xpj | 0.7 | 2.3 | 12.7 | 10.7 | 2.0 | 3.0 | 11.3 | 10.0 |
| 2zah | 16.3 | 4.7 | 17.7 | 22.3 | 2.0 | 0.0 | 32.0 | 26.3 |
| 2ztn | 0.0 | 0.0 | 34.0 | 40.0 | 2.0 | 0.0 | 32.0 | 40.0 |
| 3cji | 1.2 | 3.5 | 15.5 | 19.0 | 0.8 | 2.5 | 16.0 | 20.0 |
| 3hag | 1.0 | 3.0 | 42.0 | 42.0 | 0.0 | 2.0 | 43.0 | 43.0 |
| 3j1p | 2.0 | 5.0 | 18.3 | 36.3 | 2.0 | 4.0 | 18.3 | 37.3 |
| 3j27 | 1.7 | 1.3 | 29.8 | 26.3 | 0.5 | 0.0 | 30.2 | 27.7 |
| 3j7l | 10.0 | 1.0 | 15.3 | 17.3 | 7.0 | 0.0 | 18.3 | 18.3 |
| 3jb4 | 4.3 | 6.3 | 18.0 | 22.3 | 4.0 | 6.0 | 18.7 | 22.7 |
| 3jb8 | 12.0 | 1.0 | 13.0 | 10.0 | 12.0 | 1.0 | 13.0 | 10.0 |
| 3m8l | 0.3 | 3.3 | 44.0 | 52.7 | 0.0 | 1.0 | 44.3 | 55.0 |
| 3nap | 1.0 | 1.0 | 17.7 | 26.0 | 1.3 | 1.3 | 17.3 | 25.7 |
| 3vbs | 1.8 | 2.8 | 11.8 | 13.8 | 0.5 | 2.0 | 13.0 | 14.2 |
| 3vbu | 0.7 | 1.3 | 15.0 | 18.3 | 0.3 | 0.7 | 15.3 | 19.0 |
| 3zx8 | 19.0 | 10.0 | 22.3 | 23.7 | 15.0 | 6.0 | 26.3 | 27.3 |
| 3zx9 | 19.0 | 10.0 | 25.0 | 27.3 | 15.0 | 6.0 | 29.0 | 31.3 |
| 4aed | 2.2 | 3.5 | 11.5 | 14.2 | 1.0 | 2.0 | 12.8 | 15.5 |
| 4biq | 0.7 | 1.7 | 15.3 | 22.7 | 0.0 | 1.0 | 16.0 | 23.3 |
| 4fsj | 18.0 | 0.0 | 12.0 | 15.7 | 17.0 | 0.0 | 13.0 | 15.7 |
| 4ftb | 18.0 | 0.0 | 12.0 | 16.0 | 17.0 | 0.0 | 13.0 | 16.0 |
| 4gh4 | 1.0 | 3.0 | 16.8 | 15.8 | 0.8 | 1.5 | 17.0 | 17.2 |
| 4llf | 16.3 | 4.0 | 14.5 | 17.3 | 2.0 | 0.0 | 28.8 | 21.3 |
| 4nia | 5.0 | 1.0 | 10.0 | 7.0 | 5.0 | 1.0 | 10.0 | 7.0 |
| 4nwv | 13.0 | 0.0 | 29.7 | 34.7 | 12.0 | 0.0 | 30.7 | 34.7 |
| 4nww | 0.0 | 0.0 | 35.7 | 35.0 | 0.0 | 0.0 | 35.7 | 35.0 |
| 4oq8 | 7.0 | 2.0 | 13.0 | 10.0 | 5.0 | 1.0 | 15.0 | 11.0 |
| 4qpg | 2.0 | 2.3 | 18.7 | 18.7 | 0.7 | 0.7 | 20.0 | 20.3 |
| 4qpi | 2.7 | 4.0 | 18.7 | 20.3 | 0.7 | 3.0 | 20.7 | 21.3 |
| 4sbv | 14.7 | 0.7 | 11.3 | 9.0 | 12.0 | 0.0 | 14.0 | 9.3 |
| 4wm8 | 1.8 | 3.0 | 13.2 | 14.2 | 0.8 | 1.0 | 14.2 | 16.2 |
| 4z92 | 4.0 | 5.0 | 17.0 | 22.7 | 3.0 | 1.7 | 18.0 | 26.0 |
| 4zor | 0.0 | 0.0 | 10.0 | 12.8 | 0.0 | 0.0 | 10.0 | 12.8 |
| 5a32 | 0.0 | 2.0 | 21.0 | 25.5 | 0.0 | 1.0 | 21.0 | 26.5 |
| 5a8f | 0.7 | 1.0 | 17.0 | 13.7 | 0.3 | 0.7 | 17.3 | 13.7 |
| 5aoo | 1.0 | 2.3 | 13.0 | 18.3 | 5.3 | 6.3 | 8.7 | 14.3 |
| 5c4w | 2.0 | 3.8 | 11.0 | 13.5 | 0.5 | 1.5 | 12.5 | 15.5 |
| 5cdc | 2.7 | 4.7 | 14.3 | 24.7 | 2.0 | 3.0 | 15.0 | 26.3 |
| 5cfc | 0.8 | 1.8 | 13.2 | 13.0 | 1.0 | 2.7 | 17.7 | 15.7 |
| 5g52 | 1.3 | 2.7 | 20.7 | 25.0 | 1.0 | 4.0 | 21.0 | 23.7 |
| 5gka | 5.3 | 6.7 | 9.0 | 15.0 | 5.0 | 6.3 | 9.3 | 15.3 |
| 5ire | 0.5 | 0.0 | 29.0 | 25.3 | 2.0 | 0.0 | 28.0 | 25.3 |
| 5iz7 | 1.7 | 0.7 | 29.0 | 24.7 | 2.0 | 0.0 | 28.5 | 25.0 |

Table S5. Results for the total number $N$ of residues carrying positive ($N^+$) or negative ($N^-$) charge on the N-tails ($N_t$) and CPs ($N_c$) of viruses in the dataset (cont'd).

| **PDB ID** | STRIDE: $N_t^+$ | $N_t^-$ | $N_c^+$ | $N_c^-$ | MD2: $N_t^+$ | $N_t^-$ | $N_c^+$ | $N_c^-$ |
|---|---|---|---|---|---|---|---|---|
| 5j96 | 2.0 | 2.7 | 20.7 | 31.3 | 1.3 | 2.3 | 21.3 | 31.7 |
| 5j98 | 1.3 | 2.3 | 22.0 | 32.0 | 1.3 | 2.3 | 22.0 | 32.3 |
| 5jzg | 6.0 | 9.0 | 20.3 | 17.0 | 1.0 | 1.7 | 25.3 | 24.3 |
| 5k0u | 4.0 | 5.0 | 15.0 | 14.0 | 0.8 | 1.5 | 18.2 | 17.5 |
| 5l7o | 5.0 | 6.7 | 15.3 | 22.7 | 1.3 | 1.3 | 19.0 | 28.0 |
| 5lk7 | 1.7 | 2.7 | 23.3 | 34.7 | 1.3 | 2.3 | 23.7 | 35.0 |
| 5lk8 | 2.3 | 4.3 | 24.7 | 35.7 | 1.3 | 2.3 | 25.7 | 37.7 |
| 5lvc | 1.7 | 4.0 | 9.0 | 16.0 | 5.3 | 6.3 | 8.0 | 15.0 |
| 5lwg | 3.5 | 4.8 | 11.0 | 17.5 | 1.8 | 2.2 | 12.5 | 20.0 |
| 5lwi | 1.3 | 1.7 | 17.3 | 27.0 | 1.7 | 2.3 | 17.0 | 26.3 |
| 5mqc | 1.0 | 2.0 | 19.0 | 27.0 | 0.7 | 2.3 | 19.7 | 26.7 |
| 5ned | 1.8 | 3.5 | 17.0 | 16.8 | 0.5 | 2.2 | 18.2 | 18.0 |
| 5u4w | 0.2 | 1.0 | 16.3 | 14.8 | 1.8 | 0.8 | 16.4 | 18.0 |
| 5wsn | 1.0 | 1.0 | 28.5 | 22.7 | 0.5 | 0.5 | 29.0 | 23.2 |
| 5wte | 3.3 | 5.0 | 18.7 | 19.7 | 0.7 | 3.0 | 21.3 | 21.7 |
| 5wtf | 2.0 | 2.3 | 19.0 | 19.0 | 0.7 | 0.7 | 20.3 | 20.7 |

Table S5. Results for the total number $N$ of residues carrying positive ($N^+$) or negative ($N^-$) charge on the N-tails ($N_t$) and CPs ($N_c$) of viruses in the dataset (cont'd).

[1] Berman H M, Westbrook J, Feng Z, Gilliland G, Bhat T N, Weissig H, Shindyalov I N and Bourne P E 2000 *Nucleic Acids Res.* **28** 235–242

[2] Carrillo-Tripp M, Shepherd C M, Borelli I A, Venkataraman S, Lander G, Natarajan P, Johnson J E, Brooks C L and Reddy V S 2009 *Nucleic Acids Res.* **37** D436–D442

[3] National Centre for Biotechnology Information (NCBI) 1988 Nucleotide Database Accessed: August 2017 URL `https://www.ncbi.nlm.nih.gov/nucleotide/`



\end{document}